\documentstyle[12pt]{article}
\begin{document}

\title{
On intrinsic structure of wave function of fermion triplet
in external monopole field}

\author {V.M.Red'kov \\
     Institute of Physics, Belarus Academy of Sciences\\
F Skoryna  Avenue 68, Minsk 72, Republic of Belarus\\
e-mail: redkov@dragon.bas-net.by}

\maketitle

\begin{abstract}

Using the Weyl-Tetrode-Fock tetrad formalism, the problem of the fermion
triplet in the external monopole field ('t Hooft-Polyakov's) is
examined all over again. Spherical solutions  corresponding  to  the
total  conserved momentum $J_{i}= ( l_{i} +  S_{i} + T_{i})$, are
constructed.  The $(\theta ,\phi )$-dependence  is expressed  in
terms  of  the~Wigner's  functions
$D^{j}_{-m,\pm 1/2}(\phi ,\theta ,0)$ and
$D^{j} _{-m,\pm 3/2} (\phi ,\theta ,0)$.
The~radial
system  of  12  equations decomposes  into two sub-systems (each of
them has 6 equations) by diagonalizing some  complicated inversion
operator (acting on Lorentzian and isotopic coordinates:
$\hat{N}_{S.}= (\; \hat{\pi} \otimes  \hat{\Pi}_{bisp.}\; ) \otimes
\hat{P}$.  The case of minimal $j = 1/2$ is considered separately.
Further and more detailed analysis  has been accomplished for  the  case
of simplest external monopole-like field, namely, the one produced by putting
the Dirac monopole potential into the non-Abelian  scheme.  Now,  a  discrete
operation being able to be diagonalized,  contains  an  additional  complex
parameter $A$ :
$\hat{N}^{A}_{S.}= ( \; \hat{\pi} _{A} \otimes
\hat{\Pi}_{bisp.} \; ) \otimes  \hat{P}$.
The  same  quantity  enters  (as a parameter) basic wave functions.
Evidently, this quantity can manifest itself
at calculating matrix elements of some physical observables
$<\Psi ^{A}_{jm\delta } \mid \hat{G} \mid \Psi ^{A}_{j'm'\delta'}>$.
In particular, there have been  analyzed the $N_{A}$-parity  selection
rules arisen , those  depending on  this  $A$-parameter  explicitly.
As  shown  in  the~paper, such an $A$-freedom pointed  is a specific
consequence  of  that  there exists additional matrix operations
(symmetry of the relevant Hamiltonian). In the Cartesian isotopic
gauge, the discrete $N_{A}$-operation depends on space coordinates
explicitly:
$$
\hat{N}^{A}_{Cart.} =  \left (\; \exp (\; iA\;
\vec{t} \;\vec{n}_{\theta ,\phi }\; )  \;\hat{\Pi}_{vect.}\right )
\otimes \hat{\Pi}_{bisp.} \otimes \hat{P} \; .
$$

\noindent The wave functions considered exhibit  else one kind  of
freedom  (named  as $B$-freedom and  associated in turn  with its own
symmetry  transformations acting on the Hamiltonian).
The entering of those $A$  and $B$  parameters  looks
most neatly in the Schwinger unitary gauge: ($\delta  = \pm  1$):
$$
\Psi ^{A,B}_{jm\delta }(x)  =  \left [\;  T_{+1} \otimes \Phi ^{+}_{jm}(x)\;
 + \; e^{iB} \; T_{0} \otimes \Phi ^{0}_{jm}(x)  \; + \;
e^{iA}  \; \delta \; T_{-1} \otimes \Phi ^{-}_{jm}(x)\; \right ]
$$

\noindent There has been examined the form of the transformation  introduced:
$U(A'' \rightarrow A')$ and $U(B'' \rightarrow  B')$,
relating all possible basic wave functions associated
with different values of $A$ and $B$; their explicit forms are calculated
in  the~unitary and Cartesian gauge.

\end{abstract}

\newpage
\subsection*{1. Introduction}

The puzzle of monopole seems to be one of still yet unsolved problems
of particle physics. Apparently, together with the search of new
decided experiments and solutions of some nonlinear systems of equations,
the task of analyzing  already established results is worth attention too.
So,  the  basic frame of the present
investigation will be analysis  of particle isotopic triplet with
Lorentz spin $S=1/2$ (earlier a similar approach was developed for
the doublet case [1]; the triplet case looks in some aspect like the
doublet one except there exist quite noticeable distinctions
resulting from other isotopic structure of the triplet case which
provides some novel physical features.

Some remarks on techniques used below might be of help ; let us give
them.  The innovations of the present treatment consists in utilizing,
instead of the so-called monopole harmonics formalism [2-11], the
most conventional  Wigner's $D$-function techniques [12-19], and what
is more, in applying  the generally relativistic tetrad formalism of
Tetrode-Weyl-Fock-Ivanenko (TWFI) [20-29]. Unification through the use of
the generalized Schr\"{o}dinger's basis [29]   will apparently
simplify the real calculations carried out. In addition, the latter
makes possible to reveal  connection between the monopole topics and
the Pauli's investigation [30] concerning the problem of allowable
spherically symmetric  wave functions in quantum mechanics; his results
bear on the Dirac's $eg$-quantization condition ($e$ and $g$ are
respectively an electric and magnetic charge).

After those opening and general statements, some more particular
remarks referring to the present study and delineating its contains
are to be given.
In Sec. 1, a starting form of the relevant wave equation is specified by
choosing particular bases  simultaneously in the tetrad and isotopic
frames; those are respectively the spherical tetrad basis and
isotopic  Schwinger's (about terminology used  see in Supplement A).
The radial equations found by separation of variables are
rather complicated. We simplify them by searching a suitable operator
which could be diagonalized simultaneously  with the
 $\vec{j}^{2}$ and $j_{3}$. As well known, the usual space reflection
($P$-reflection) operator for a bispinor field has to be followed by
certain transformation in the isotopic space so that a
required quantity could be
constructed. However, the solution of this problem known to date is
not general as much as possible: some possibilities have remained
unused. For this reason the question of reflection symmetry  [31-59]
in the particle triplet-monopole system is examined here in full detail. As
a result we  find that there exist two different possibilities
depending on what type of external monopole potential is  analyzed.
So, in case of the simplest one  [60]   that can be regarded as result of
embedding the Abelian potential into the non-Abelian scheme
the composite  reflection operator $\hat{N}_{A}$ is determined
ambiguously: it depends on an arbitrary  complex numerical  parameter
$A , (e^{iA} \neq 0 )$. In turn, if we treat the non-trivial
monopole  case [61-63]),  this
$e^{iA}$ must be equated to $1$.

Else one problem deserving to be mentioned
concerns distinctions between the Abelian and non-Abelian monopoles.
We draw attention to the fact that, in the non-Abelian case, two
systems: a free triplet and a triple affected by monopole potential
(wether a trivial or not-trivial one), they   have their spherical
symmetry operators  $\vec{j}^{2}, \; j_{3}$ identically alike.
Correspondingly, in both  these cases, isotriplet wave functions do
not vary at all in their dependence on angular variables $\theta,
\phi$ . This  non-Abelian wave functions' property sharply
contrasts with the Abelian one where both electronic wave functions and
corresponding symmetry operators undergo
significant transformations   in the external monopole field
($D^{j}_{mm'}$ designates the Wigner's $D$-functions;
the value $eg = 0$ relates to the free electronic case):
$$
J^{eg}_{1}= l_{1} + { (i \sigma ^{12} - eg) \cos \phi \over
\sin\theta } \; , \qquad
J^{eg}_{2}= l_{2} + { (i \sigma ^{12} - eg) \sin \phi \over
\sin\theta } \; , \qquad J^{eg}_{3} = l_{3}
\eqno(1.1)
$$
$$
\Phi^{eg}_{jm}(t,r,\theta ,\phi ) = {e^{-i\epsilon t}\over r}\;
\left ( \begin{array}{l}
          f_{1}(r) \; D^{j}_{-m,eg-1/2}(\phi ,\theta ,0) \\
          f_{2}(r) \; D^{j}_{-m,eg+1/2}(\phi ,\theta ,0) \\
          f_{3}(r) \; D^{j}_{-m,eg-1/2}(\phi ,\theta ,0) \\
          f_{4}(r) \; D^{j}_{-m,eg+1/2}(\phi ,\theta ,0)
\end{array} \right )
\eqno(1.2)
$$

\noindent In other words, we may state that one of the fundamental
features underlying the theory of the non-Abelian monopole is that
such a field does not destroy (or does not touch) the angular
dependence which is dictated  solely by isotopic structure
arguments. In that sense, it represents an analogue of a spherically
symmetric Abelian potential $A_{\mu}(x) = (A_{0}(r), 0,0,0)$ rather
than  the Abelian  monopole potential  $A_{\mu}(x) = (0,0,0, A_{\phi}
= g \cos \theta )$. In that connection we might draw attention to the
fact that the designation itself {\it monopole} anticipates interpretation
of the vector  triplet $A_{\mu}^{(a)}(x)$  as carrying, in a new situation,
the  old Abelian monopole quality and essence, although a real  degree of
their similarity   would be much less that one might expect.

In Sec. 3, we proceed further with the reflection symmetry and
give some attention to  the question of explicit form of the above
operator $\hat{N}^{S.}_{\alpha}$ (here the sign $S.$ stands for the
Schwinger's isotopic gauge) in some other
isotopic gauges. There is no  reason {\em a priori}   to
choose a unique gauge as more preferred that all others; the
particular choice above was  dictated  by convenience only. That
restriction can easily be relaxed, for example, by demonstration of
several more gauges. We are chiefly interested in the Dirac unitary
and Cartesian gauges (as the most frequently used in the literature).

In Sec. 4,
because the matter of discrete (in particular, $P$-inversion) always
gets much attention on many physical reasons (in non-Abelian as
well as Abelian theories)  and because the freedom in choosing a
discrete operator $\hat{N}_{a}$ seems at first glance rather unusual
and even puzzled,  we proceed with further studying monopole-affected
manifestations of that symmetry.
In the Abelian case, because of the well known monopole $P$-violation,
electronic wave function in monopole field do not obey a fundamental
structural condition
$$
\Psi^{eg}_{\epsilon j m \delta}(t, -\vec{r}) =
(\; matrix \;) \;
\Psi^{eg}_{\epsilon j m \delta}(t, +\vec{r})
\eqno(1.3a)
$$

\noindent  which does guarantee the existence of the parity
selection rules. Instead, only the following
$$
\Psi^{eg}_{\epsilon j m \delta}(t, -\vec{r}) =
(\;matrix\;) \;
\Psi^{-eg}_{\epsilon j m \delta}(t, +\vec{r})
\eqno(1.3b)
$$

\noindent holds. It should be noted that in the literature there have
been several suggestions as to how obtain a certain  (formal)
covariance of the monopole situation with respect to the
$P$-symmetry. A single actual  outcome of such attempts is that
all those  really imply the pseudoscalar character (under
$P$-inversion) of a  magnetic charge. To avoid misunderstanding, one
thing should be noticed: the use of various gauges at description of
the monopole (Dirac, Schwinger, Wu-Yang's and so on) brings some
peculiarities  to  this.  Indeed, in
Schwinger representation,  a  pseudo-scalar nature of a  magnetic charge
has usual sense, whereas the same situation in two other gauges
seems like as if the common $P$-operation needs to be additionally
improved through slight alterations (those latter can be found from
the involved gauge transformations). But, admittedly,
pseudoscalar-based suggestions do not permit to overcome the
non-existence of the discrete symmetry selection rules for matrix
elements at considering one particle problem in a fixed monopole
potential.  In contrast to the Abelian monopole case, the
non-Abelian situation is quite different: a relation with the
required structure there exists (at first,  just the case
$\hat{N}_{A=0}$ -symmetry is analyzed)
$$
\Psi^{triplet}_{\epsilon j m \delta}(t, -\vec{r}) =
( \; matrix \; ) \;
\Psi^{triplet}_{\epsilon j m \delta}(t, +\vec{r})
\eqno(1.4)
$$

\noindent correspondingly   $N_{A=0}$-parity selection rules may be
produced; we give some details on this matter.
We consider the question of how the complex parameter
$A$ being involved into  $\hat{N}_{a}$-operator and
corresponding triplet wave functions can manifest itself in matrix
elements. To this end, an explicit expression for possible matrix
elements is looked into. As a natural and simple illustration, the
above problem of parity selection rules is investigated again, but
now depending on that $A$-background. Taking in mind the class of
composite physical observables, that in general may  have some
inclusive constituent structure, a new definition of
composite scalars and pseudoscalars with respect to the above
$\hat{N}_{A}$-operation naturally occurs. At this, different values
of $A$ will lead to different concepts of scalars and  pseudoscalars.
Correspondingly, $N_{A}$-parity selection rules arising in sequel for
matrix elements (certainly if an observable belongs to the class of
those $N_{A}$-scalars or pseudoscalars) differ basically from each
other.

In Sec. 5 we are especially interested in the question: where does
the above ambiguity come from? It is quite easily understandable that
this possibility is closely connected with the fact of decoupling of
three isotopic components in the wave equation
itself, so that one may change independently three isotopic
amplitudes and in the same time not destroy already given
$j,m$-structure, only touching the $N_{A}$-structure.
In more physically oriented studies, that possibility
may be thought of in terms of electric charge's characteristics of
different isotopic components; namely, together with $\hat{j}^{2}$
and $j_{3}$, the third component of isotopic spin $t_{3}$ (electric
charge operator) may be diagonalized upon the wave functions
$\Psi_{\epsilon j m }$. In addition, the situation can be thought of in terms of
a hidden symmetry: there are two operators, $t_{3}$ and $\hat{N}_{A}$,
commuting with the Hamiltonian but not commuting with each other.
For the present paper's purposes just
it is an~additional symmetry-based formal approach  that seems more
appropriate and really being  made of use.  So, just on those aspects
of the problem we are going to concentrate further analysis. Thus,
it is noted that every
particular value  $A$ merely governs basis states $
\Psi^{A}_{\epsilon j m \delta}(x)$ with no change in the whole
functional space; connection between different bases
$$
\Psi^{A'}_{\epsilon j m \delta}(x)   =
V(A ',\; A)  \Psi^{A}_{\epsilon j m \delta}(x)
$$

\noindent can be factorized as follows (for more detail on the used
designations see in  Sec. 5)
$$
V(A ',\; A)   = e^{+i{A'-A \over 2}} \; D(A' - A)
\; \Delta (A' - A ) \; ,
\eqno(1.5a)
$$

\noindent   Two of these operations  vary in their acting on
particular components of the triplet wave functions  and also
vary in their affecting the $\hat{N}_{A}$-operator
$$
\Delta \; \hat{N}_{A}\; \Delta^{-1}  \;
= \;\hat{N}_{A'} \;\;\; but \;\;   \;
D \; \hat{N}_{A}\; D^{-1}  \;= \; \hat{N}_{A}
\eqno(1.5b)
$$

\noindent but both of two do not change the composite momentum
components. The second relation in (1.5b) implies that, though such a
$D$-symmetry can potentially prove itself in explicit expressions for
matrix elements (analogously the $A$-symmetry), this cannot happen to
the $N$-parity selection rules because the $D$-symmetry does not
affect the $\hat{N}_{A}$-operator.

The acting of $D$- and $\Delta$-transformations may be
spelled out in a straightforward way.  Let a general wave
function for the triplet  be
$$
\Psi^{A(B)}_{\epsilon
j m \delta}(x) = [\; T_{+1} \otimes \Phi^{(+)}(x) + e^{iB} \;
T_{0}\otimes \Phi^{(0)}(x) + \delta e^{iA} \; T_{-1} \otimes
\Phi^{(-)}(x) \; ]
\eqno(1.6a)
$$

\noindent then the  operations $\Delta(\Gamma)$ and $D(\Gamma)$
will act on those functions (1.6a) according to
$$
\Psi^{A'(B)}_{\epsilon j m \delta}(x) =   \Delta(A'-A )
\; \Psi^{A(B)}_{\epsilon j m \delta}(x) \; , \;\;
\Psi^{A(B')}_{\epsilon j m \delta}(x) =   D(B'-B) \;
   \Psi^{A(B)}_{\epsilon j m \delta}(x) \;
\eqno(1.6b)
$$

\noindent with no connections between $(B'-B)$ and $(A'-A)$:
those perimeters are  completely independent.

The operation $\Delta(A'-A)$ in itself represents
one-parametric rotation belonging to the complex 3-dimensional group
$SO(3.C)$; the $D(B'-B)$ in turn  is  a certain `conformal'
transformation in  isotopic  space. As may be thought, useful and
somewhat intriguing are their respective forms in Cartesian isotopic
gauges, which are calculated. It is explicit dependence on angular
coordinate $\theta, \phi$ that makes them so exciting and
not trivial in appearance:
$$
\Delta^{Cart.} (A'-A) = exp [\; i{A'-A \over 2} \;\vec{t} \;
\vec{n}_{\theta \phi} \; ]  \; ,  \;\;\;
D^{Cart.}(\Gamma) = [\; I \;+ \;(\; e^{-i\Gamma }\; -\; I\; )
\tilde{t}_{0} \; ]
\eqno(1.6c)
$$

\noindent  where  $\tilde{t}_{0}$ is a $\theta, \phi$-dependent
matrix.

\begin{quotation}

{\it It  may be repeated  else one time and stressed that both
these symmetry operations occur only if the case of special monopole
potential is at hand; instead, for the 't Hooft-Polyakov potential as
well as for the free isotopic triplet case no such additional
symmetries (tied with the operators $\vec{j}^{2}, j_{3}$) will arise.
This latter seems quite natural in the light of the genuinely
non-Abelian monopole presence and less natural  or even somewhat
strange as being concerned to the free triplet system. Indeed, this
mean that we have come to be able to diagonalize   an electric charge
operator in the special monopole field and cannot do so in the other
(monopole free) situation. In any case, just this shows that the case
of triplet in the simplest nonopole potential exhibits quite definite
features which set it apart from  the other  systems.
}
\end{quotation}

\subsection*{2. Separation of variables and discrete  operator}

The basic tetrad-based equation  (see in Supplement A)
$$
\left[\; i \gamma ^{\alpha }(x)\; (\; \partial_{\alpha } \; + \;
\Gamma _{\alpha }(x) \; - \;i e \; t^{a}\; W^{a}_{\alpha }(x)\; ) \; - \;
( m \; + \; \kappa \; \Phi ^{a}(x) \;t^{a} \; ) \; \right ]\; \Psi (x) = 0
\;
\eqno(2.1)
$$

\noindent  at the use of Schwinger unitary gauge in isotopic space
and the spherical tetrad basis, takes the form
$$
\left [ \; \gamma ^{0} \; (\; i \partial _{t} \; + \; e r F(r)\; t^{3} ) +
i \gamma ^{3} \; ( \partial _{r} + {1 \over r }) \;  + \; {1 \over r} \;
\Sigma _{\theta ,\phi } \; + \right.
$$
$$
 \left. + {1 \over r} \; (e r^{2} K(r) + 1) \;  (\gamma ^{1} \otimes  t^{2} -
\gamma ^{2} \otimes  t^{1}) \; -  \;
(m \; + \; \kappa  r \Phi (r) t^{3}) \;  \right ] \; \Psi ^{S.} = 0
\; ,
\eqno(2.2a)
$$

\noindent where
$$
\Sigma _{\theta ,\phi } = \left [ \; i \gamma ^{1} \; \partial _{\theta }  \;
 + \; \gamma ^{2} \;
{i \partial _{\phi } \; + \;  (i\sigma ^{12} + t^{3}) \cos \theta   \over
 \sin \theta } \;  \right ]
\; .
\eqno(2.2b)
$$

\noindent The Dirac matrices $\gamma^{\alpha}$ are chosen in the Weyl
spinor representation; the isotopic ones  $t^{a}$ are specified in
the so-called cyclic basis
$$
t^{1} = {1 \over \sqrt{2}}
\pmatrix{0 & 1 & 0 \cr 1 & 0 & 1 \cr 0 & 1 & 0 } \; ,  \qquad
t^{2} = {1 \over \sqrt{2}} \pmatrix{0 & -i & 0 \cr i & 0 & -i \cr 0 & i & 0} \; ,
\qquad t^{3} = \pmatrix{1 & 0 & 0 \cr 0 & 0 & 0 \cr 0 & 0 & -1} .
$$

Having giagonalized the operators $\hat{j}^{2}$ and  $j_{3}$, the
wave functions with quantum numbers  $j,m$ are constructed as
$$
\Psi^{S.}_{\epsilon jm}(t,r,\theta ,\phi ) =  {1 \over r} \;
e^{-i\epsilon t}   \; \times
\eqno(2.3)
$$ $$
\left [\; T_{+1} \otimes \left (
\begin{array}{c}
f_{1} D_{-3/2} \\ f_{2} D_{-1/2} \\ f_{3} D_{-3/2} \\ f_{4} D_{-1/2}
\end{array} \right )    \;   +   \;
 T_{0} \otimes   \left (  \begin{array}{c}
 h_{1}D_{-1/2} \\ h_{2}D_{+1/2} \\ h_{3}D_{-1/2} \\ h_{4}D_{+1/2}
\end{array}  \right  ) \; + \;
 T_{-1} \otimes     \left  ( \begin{array}{c}
g_{1}D_{+1/2} \\ g_{2} D_{+3/2} \\ g_{3} D_{+1/2} \\ g_{4} D_{+3/2}
\end{array}  \right ) \; \right ]
$$

\noindent where  $D_{\sigma }$ denotes the Wigner functions
$D^{j}_{-m,\sigma }(\phi ,\theta ,0)$ ( $j = 1/2, 3/2, \ldots$) ;
$f_{i}, \; g_{i}, \; h_{i}$  represent  radial functions; and
$T_{-1}, \; T_{+1}, \; T_{0} $ designate the basic vector of isotopic
space  ( $t^{3} \; T_{s} = \; s \; T_{s}$):
$$
T_{+1} = \pmatrix{1 \cr 0 \cr 0} \;  , \qquad
T_{0}  = \pmatrix{ 0 \cr 1 \cr 0} \; , \qquad
T_{-1} = \pmatrix{0 \cr 0 \cr 1} \; .
$$

The first step is to separate the variables and work out a radial
system\footnote{At this certainly an old result will be reproduced
again, though the working method itself is a new one.}.  To this end,
having used the known  relations for Wigner's $D$-functions
[18]
$$
\partial _{\theta } D_{+1/2} = {1 \over 2} ( a D_{-1/2}
- b D_{ + 3/2}) \; ,
$$
$$
[ \sin^{-1} \theta \; (i\partial _{\phi } -
(1/2)\cos \theta ) ] D_{+1/2} = {1 \over 2} (- a D_{-1/2} - b
D_{+3/2}) \; ;
$$
$$
\partial _{\theta } D_{-1/2} = {1 \over 2} ( b
D_{-3/2} - a D_{+1/2}) \; ,
$$
$$
[ \sin ^{-1} \theta \; (i\partial
_{\phi } + (1/2)\cos \theta ) ] D_{ -1/2} = {1 \over 2} ( - b
D_{-3/2} - a D_{+1/2}) \; ;
$$
$$
\partial _{\theta } D_{+3/2} = {1
\over} ( b D_{+1/2} - c D_{+5/2}) \; ,
$$
$$
 [ \sin ^{-1} \theta \;
(i\partial _{\phi }  - (3/2)\cos \theta ) ] D_{+3/2} = {1 \over 2} (-
b D_{+1/2} - b D_{+5/2}) \; ;
$$
$$
\partial _{\theta } D_{-3/2} =
1/2 ( c D_{-5/2} - b D_{-1/2}) \; ,
$$
$$
[\sin ^{-1} \theta \;
(i\partial _{\phi } + (3/2)\cos \theta ) ] D_{-3/2} = {1 \over 2} ( -
c D_{-5/2} - b D_{-1/2})
\eqno(2.4a)
$$

\noindent where
$$
a = (j + 1/2) \; , \qquad b = \sqrt{(j - 1/2)(j + 3/2)} \; ,  \qquad
c = \sqrt{(j-3/2)(j+5/2)}
$$

\noindent we find the action of
$\Sigma _{\theta ,\phi }$ on  $\Psi ^{S.}_{\epsilon jm}(x)$:
$$
\Sigma _{\theta ,\phi } \; \Psi ^{è.}_{\epsilon jm}(x)  =   \;
{1 \over r} \; e^{-i\epsilon t}  \left [\;
i\; b \; \; T_{+1}  \otimes  \left( \begin{array}{c}
              -f_{4} \; D_{-3/2} \\  +f_{3} \; D_{-1/2} \\
             + f_{2} \; D_{-3/2} \\  -f_{1} \; D_{-1/2}
\end{array} \right )  \; +  \right.
$$
$$
\left. +\; i \; a \; T_{0} \otimes
\left ( \begin{array}{c}
   - h_{4} \; D_{-1/2}  \\  + h_{3} \; D_{+1/2} \\
   + h_{2} \; D_{-1/2}  \\  - h_{1} \; D_{+1/2}
\end{array} \right )  \; + \; i \; b \; T_{-1} \otimes
\left  (  \begin{array}{c}_
 -g_{4} \; D_{+1/2} \\   + g_{3} \; D_{+3/2}   \\
+ g_{2} \; D_{+1/2} \\   - g_{1} \; D_{+3/2}
\end{array} \right ) \;     \right ] \; .
\eqno(2.4b)
$$

\noindent In addition, as a simple matter of calculation we get
the action of a mixing term ($ W \equiv (er^{2}K(r) + 1) $):
$$
{W \over r}\; ( \gamma ^{1} \otimes  t^{2}  \; - \;
\gamma ^{2} \otimes  t^{1}) \;
\Psi ^{S.}_{\epsilon jm}(x) \; = \;
{ e^{-i\epsilon t} \over r}  \sqrt{2}  \;  {W \over r} \; \times
$$
$$
\left [ \; T_{+1} \otimes \left ( \begin{array}{c}
0 \\ i h_{3} \; D_{-1/2} \\ 0 \\  -i h_{1} \; D_{-1/2}
\end{array} \right ) \; + \; T_{0} \otimes
\left ( \begin{array}{c}
 -i f_{4} \; D_{-1/2} \\  + i g_{3} \; D_{+1/2} \\
 +i f_{2} \; D_{-1/2} \\   -i g_{1} \; D_{+1/2}
\end{array} \right ) \; + \; T_{-1}  \otimes
 \left  ( \begin{array}{c}
    - i h_{4} \; D_{+1/2} \\ 0 \\ + i h_{2} \; D_{+1/2} \\ 0
\end{array} \right ) \right ]
\eqno(2.4c)
$$

\noindent one can note that three  distinct isotopic components
mingle just through the terms which are proportional to the Wigner's
functions  $D_{\pm 1/2}$.  Finally, after simple calculating we
shal find the following set of radial equations (for saving space,
the notation $ \tilde{F} = e r F(r) , \tilde{\Phi } = \kappa r \Phi
(r)$ is used):
$$
(\epsilon  + \tilde{F}) \; f_{3} \; -  \; i {d \over dr}\;  f_{3} \; - \;
{ib \over r} \; f_{4} \;  - \; (m + \tilde{\Phi })\;  f_{1} = 0   \; ,
$$
$$
(\epsilon  + \tilde{F}) \;  f_{4} \; + \; i {d \over dr}\;  f_{4} \; +\;
 {ib \over r}\;  f_{3}\;  + \;
i {\sqrt{2} W \over r}  h_{3} \;-\; (m + \tilde{\Phi })\; f_{2} = 0   \; ,
$$
$$
(\epsilon  + \tilde{F}) \; f_{1} \;
+ \; i{d \over dr} \; f_{1} \; + \; {ib \over r}\;  f_{2} \;  - \;
 ( m + \tilde{\Phi }) f_{3} = 0   \; ,
$$
$$
(\epsilon   + \tilde{F}) \;  f_{2}  \; - \;
i{d \over dr} \; f_{2} \; - \; {ib \over r} \; f_{1} \;  - \;
i {\sqrt{2} W \over r} \; h_{1} \;  - \;
 (m   + \tilde{\Phi }) \; f_{4} = 0   \; ;
\eqno(2.5a)
$$
$$
(\epsilon  - \tilde{F})\; g_{3} \; - \; i {d \over dr} \; g_{3} \; -\;
 {ib \over r}\;  g_{4} \; - \;
i {\sqrt{2} W \over r}\;  h_{4} \; -\;  (m  -  \tilde{\Phi })\; g_{1} = 0  \; ,
$$
$$
(\epsilon  - \tilde{F}) \; g_{4} \; +\;  i {d \over dr} \; g_{4}\;  +\;
 {ib \over r}\;  g_{3} \; -\;
(m - \tilde{\Phi })\;  g_{2} = 0      \; ,
$$
$$
(\epsilon  - \tilde{F})\;  g_{1} \; + \; i{d \over dr}\;  g_{1}\; + \;
{ib \over r} \; g_{2} \; + \;
i  {\sqrt{2} W \over r}\;  h_{2} \;  - \;  (m - \tilde{\Phi }) g_{3} = 0 \; ,
$$
$$
(\epsilon   - \tilde{F}) \; g_{2} \; - \; i {d \over dr} \; g_{2}\;  - \;
{ib \over r}\;  g_{1}\;   - \;  (m - \tilde{\Phi }) g_{4} = 0   \; ;
\eqno(2.5b)
$$
$$
\epsilon  h_{3}\;  -\;  i{d \over dr} \; h_{3}\;  -\;  {ia \over r} \; h_{4}\;  - \;
i {\sqrt{2} W \over r} \; f_{4} \;  -  \;  m h_{1} = 0 \; ,
$$
$$
\epsilon \; h_{4} \; + \; i {d \over dr} \; h_{4} \; + \; { ia \over r}\; h_{3}
 \; + i \;  {\sqrt{2} W \over r} \;  g_{3} \; - m h_{2} = 0  \; ,
$$
$$
\epsilon \;  h_{1} \; + \; i{d \over dr}\;  h_{1} \; + \; {ia \over r}\; h_{2} \; + \;
i {\sqrt{2} W \over r} \; f_{2}\;  - m \;  h_{3} = 0   \; ,
$$
$$
\epsilon\;  h_{2} \; -\;  i {d \over dr}\; h_{2} \;- \; {ia \over r}\; h_{1}\;  -\;
i {\sqrt{2} W \over r}\;  g_{1}\;  -\;  m h_{4} = 0    \; .
\eqno(2.5c)
$$

Now  let us try to simplify these equations (2.6) by diagonalizing
additionally a suitable discrete operator: a composite reflection in
the Lorentzian  and  isotopic spaces.
The usual bispinor $P$-inversion has, in the used tetrad basis, the
form
$$
\hat{\Pi}_{sph.}  = \left  ( \begin{array}{cccc}
       0 & 0 & 0 &  -1 \\  0 & 0 & -1 & 0 \\
       0 &  -1 & 0 & 0 \\ -1 & 0 &  0 & 0
\end{array} \right ) \otimes \hat{P}  =  \;
( - \gamma ^{5} \; \gamma ^{1} ) \otimes  \hat{P} \;
\eqno(2.6a)
$$

\noindent the $\hat{P}$ is determined by relation
$\hat{P} \Psi (\theta ,\phi ) =
\Psi (\pi - \theta , \phi + \pi )$).
Correspondingly, that $\hat{\Pi}_{sph.}$ acts upon the composite
wave function   as follows (the relationship
$ \hat{P} \; D^{j}_{-m,\sigma } =
(-1)^{j} \; D^{j}_{-m,-\sigma}$  is taken into account)
$$
\hat{\Pi}_{sph.}\; \Psi _{\epsilon jm} ^{S.}(t,r,\theta ,\phi )  =
{e^{-i\epsilon t}  \over r} \; (-1)^{j+1} \times
$$
$$
\left [ \;
T_{+1} \otimes   \left     ( \begin{array}{c}
f_{4} \; D_{+1/2} \\ f_{3} \; D_{+3/2} \\ f_{2} \; D_{+1/2} \\
f_{1} \; D_{+3/2}      \end{array}  \right  )       \;  + \;
T_{0} \otimes \left  ( \begin{array}{c}
h_{4} \; D_{-1/2} \\ h_{3} \; D_{+1/2}  \\ h_{2} \; D_{-1/2} \\
h_{1} \; D_{+1/2}      \end{array} \right ) \;
+    \;   T_{-1} \otimes \left   (     \begin{array}{c}
g_{4} \; D_{- 3/2} \\ g_{3} \; D_{-1/2} \\
g_{2} \; D_{-3/2} \\ g_{1} \; D_{-1/2}
\end{array} \right ) \; \right  ] \;  .
\eqno(2.6b)
$$

\noindent From this it follows immediately that the
$\hat{\Pi}_{sph.}$ cannot be diagonalized upon the functions of the
sort (2.3); but an operator with required properties can be
constructed through extending of the above $\hat{\Pi}_{sph.}$    by
a special transformation    $\hat{\pi }$
upon isotopic coordinates and taking the
unit vectors  $\{\;T_{+1}, \;  T _{0} ,\;  T_{-1}\; \}$
into   $\{ \;  T_{-1}, \; T _{0} , \; T_{+1} \; \}$
respectively (possibly apart from some number factors).
Expecting the $\hat{\pi }$  to satisfy
$$
\hat{\pi } \; T_{+1}  = \alpha \; T_{-1} \; ,  \qquad
\hat{\pi } \; T_{0}   = \beta  \; T_{0}  \; ,  \qquad
\hat{\pi } \; T_{-1}  = \gamma \; T_{+1} \; ,
\eqno(2.7a)
$$

\noindent we will find its matrix representation
$$
\hat{\pi } = \pmatrix{0 & 0 & \gamma \cr 0 & \beta & 0 \cr
\alpha & 0 & 0} \; .
\eqno(2.7b)
$$

\noindent The required composite operator  $\hat{N}$, being
determined by
$$
\hat{N} \;  =\;  ( \hat{\pi } \otimes
\hat{\Pi}_{sph.})
\eqno(2.7c)
$$

\noindent will act on the wave functions as follows
$$
\hat{N} \;  \Psi ^{S.}_{\epsilon jm}(t,r,\theta ,\phi ) =
{e^{-i\epsilon t}  \over r} \;  (-1)^{j+1} \times
$$
$$
\left  [ \;
\gamma \; T_{+1} \otimes \left  ( \begin{array}{c}
  g_{4} \; D_{-3/2} \\  g_{3} \; D_{-1/2} \\
  g_{2} \; D_{-3/2} \\  g_{1} \; D_{-1/2}
\end{array} \right )
\;    + \;  \beta \; T_{0}  \otimes  \left  ( \begin{array}{c}
h_{4} \; D_{-1/2} \\ h_{3} \; D_{+1/2} \\
h_{2} \; D_{-1/2} \\ h_{1} \; D_{+1/2}
\end{array} \right ) \; + \;  \alpha \; T_{-1} \otimes
\left ( \begin{array}{c}
f_{4} \; D_{+1/2} \\ f_{3} \; D_{+3/2} \\
f_{2} \; D_{+1/2} \\ f_{1} \; D_{+3/2}
\end{array} \right ) \; \right ]\; .
\eqno(2.8a)
$$

\noindent From the proper value equation
$\hat{N} \; \Psi ^{S.}_{\epsilon jm}  = N \; \Psi ^{S.}_{\epsilon jm}$
it follows
$$
 (-1)^{j+1}  \gamma \; g_{4} = N \; f_{1}  \;   , \qquad
 (-1)^{j +1} \alpha \; f_{1} = N \; g_{4}  \;   ,
$$
$$
 (-1)^{j+1}  \gamma \; g_{3} = N \; f_{2}  \;   ,   \qquad
 (-1)^{j+1}  \alpha \; f_{2} = N \; g_{3}  \;   ,
$$
$$
  (-1)^{j+1} \gamma \; g_{2} = N \; f_{3}  \;   ,   \qquad
  (-1)^{j+1} \alpha \; f_{3} = N \; g_{2}  \;   ,
$$
$$
  (-1)^{j+1} \gamma \; g_{1} = N \; f_{4}  \;   ,    \qquad
  (-1)^{j+1} \alpha \; f_{4} = N \; g_{1}  \;   ;
\eqno(2.8b)
$$
$$
  (-1)^{j+1} \beta  \; h_{4} = N \; h_{1}  \;   ,     \qquad
  (-1)^{j+1} \beta  \; h_{1} = N \; h_{4}  \;   ,
$$
$$
  (-1)^{j+1} \beta  \; h_{2} = N \; h_{3}  \;   ,     \qquad
  (-1)^{j+1} \beta  \; h_{3} = N \; h_{2}  \;   .
\eqno(2.8c)
$$

\noindent Then, from (2.8c) we get  $N^{2} \;=\;
(-1)^{2j+2}\; \beta ^{2}$; from (2.8b)  it follows $ N^{2} =
(-1)^{2j+2} \; (\alpha \; \gamma )$; thus
$\beta ^{2} = (\alpha \; \gamma )$. In addition, noting that
$$
\hat{\pi }^{2}  = \pmatrix{\alpha  \gamma & 0 & 0
\cr 0 & \beta^{2} & 0 \cr 0 & 0 & \alpha  \gamma }   = \beta^{2} \; I
$$

\noindent and accepting that the choice of $\beta $ is not
material to diadonalizing that discrete operator, so that we may take
$\beta  = 1$, and  eventually get $\gamma  = \alpha ^{-1})$ as a
single free numerical parameter at an ambiguous  operator
$\hat{\pi }_{\alpha }$. Thus,  we have worked out two values for
the $N_{\alpha}$-parity and concomitant with them limitations on
radial functions:
$$
N = \delta \; (-1)^{j+1} , \delta  = \pm  1
$$
$$
h_{3} = \delta \; h_{2} \; , \qquad h_{4} = \delta \; h_{1} \; , \qquad
g_{4} = \delta \; \alpha  f_{1} \; ,
$$
$$
g_{3} = \delta \; \alpha f_{2} \;  ,\qquad
g_{2} = \delta \; \alpha f_{3} \;  , \qquad
g_{1} = \delta \alpha f_{4}  \;    ;
\eqno(2.9a)
$$

\noindent so that the functions with quantum numbers  $\epsilon jm\delta$
are built as
$$
\Psi ^{S.\;(\alpha)}_{\epsilon jm\delta }(t,r,\theta ,\phi )  =
{e^{-i\epsilon t} \over r}
$$
$$
\left   [  \;  T_{+1} \otimes
\left ( \begin{array}{c}
f_{1} \; D_{-3/2} \\ f_{2} \; D_{-1/2} \\
f_{3} \; D_{-3/2} \\ f_{4} \; D_{-1/2}
\end{array} \right    )  \;   +  \; T_{0} \otimes
\left    (  \begin{array}{c}
h_{1} \; D_{-1/2} \\ h_{2} \; D_{+1/2} \\
\delta \; h_{2} \; D_{-1/2} \\ \delta \; h_{1} \; D_{+1/2}
\end{array}    \right ) \;  + \; \delta \; \alpha \; T_{-1} \otimes
\left ( \begin{array}{c}
f_{4} \; D_{+1/2} \\ f_{3} \; D_{+3/2} \\
f_{2} \; D_{+1/2} \\ f_{1} \; D_{+3/2}
\end{array}    \right ) \; \right ]     .
\eqno(2.9b)
$$

\noindent At fixed  $(\epsilon,\; j, \; m)$  the quantity
$\delta$ takes two values:  $+1$ or  $-1$; the presence of the
$\alpha $  in (2.9b)  reflects the ambiguity in choosing the discrete
$\hat{N}_{\alpha}$.

Now, we are going to substitute (2.9a) into equations (2.6), so that
together with the question of their self-consistency we  will be
able, by the same token, study the  question of the possible
commuting of the $\hat{N}_{\alpha}$ with the relevant
triplet-fermion-monopole Hamiltonian. The whole situation here seems
completely analogous to that appeared in studying the
doublet-fermion-monopole case [1].
It turns out  that the system so obtained will not be
self-consistent if the two terms in the equation (2.1) are
valid: those are $e r F(r)\; t^{3}$ and $ \kappa\; r \Phi (r)\;
t^{3}$.  So, we have to set  $F(r)=O$ as well as $\kappa = 0 $; that
is to restrict ourselves to the case of purely monopole (not dyon)
potential and of null coupling  constant  $\kappa$.    Thus,
we arrive at
$$
\epsilon  \; f_{3} \; - \;  i{d \over dr} \; f_{3} \; - \; {ib \over r}\; f_{4}
\; - \; m\; f_{1} = 0
$$$$
\epsilon\; f_{4} \;+\; i {d \over dr}\; f_{4} \;+\;  {ib \over r} \; f_{3}\;+\;
i \delta\; {\sqrt{2} W \over r} \; h_{2} \; -\;  m \; f_{2} = 0
$$$$
\epsilon \; f_{1}\; + \; i{d \over dr} \;  f_{1} \; + \;  {ib \over r} \; f_{2}
\;  -\;  m \; f_{3} = 0
$$$$
\epsilon \;  f_{2}\;  -\;  i{d \over dr} \; f_{2} \; -\; {ib \over r} \; f_{1}
\; - \; i {\sqrt{2} W \over r} \; h_{1} \; - \; m \; f_{4}  = 0
\eqno(2.10a)
$$$$
\epsilon \; f_{2} \; - \; i{d \over dr}\;  f_{2} \; - \; {ib \over r} \; f_{1}
\; - \; i \alpha ^{-1} \; {\sqrt{2} W \over r} \;  h_{1}\;  - \; m \; f_{4} = 0
$$$$
\epsilon \; f_{1} \; + \;i {d \over dr}  \; f_{1} \; + \; {ib \over r} \;
 f_{2} - m \; f_{3} = 0
$$$$
\epsilon \; f_{4} \; + \; i{d \over dr} \; f_{4} \; + \; {ib \over r} \;
 f_{3} \; + \;
i \delta \; \alpha ^{-1} \; {\sqrt{2} W \over r} \; h_{2} \; - \;  m \; f_{2} = 0
$$$$
\epsilon \; f_{3} \; - \; i {d \over dr}\; f_{3} \; - \; {ib \over r}\;
 f_{4}\; -\;  m\;  f_{1} = 0
\eqno(2.10b)
$$$$
\epsilon \; h_{2} \; - \; i{d \over dr} \; h_{2} \; -\; {ia \over r}\; h_{1}
\;  - \;  i \delta\; {\sqrt{2} W \over r}\;  f_{4} \; -
\; \delta \; m \; h_{1} = 0
$$$$
\epsilon \; h_{1} \; + \; i{d \over dr}\; h_{1} \; + \; {ia \over r}\; h_{2}
\; + \; i \alpha  \;  {\sqrt{2} W \over r} \; f_{2}\; - \;
\delta \; m \; h_{2} = 0
$$$$
\epsilon \; h_{1} \; + \; i {d \over dr} \; h_{1} \; + \; {ia \over r} \; h_{2}
\; + i \; {\sqrt{2} W \over r} \; f_{2} \;  - \delta \; m \; h_{2} = 0
$$$$
\epsilon \; h_{2} \; - \; i {d \over dr}\; h_{2} \; - \; {ia \over r} \; h_{1}
\; - \; i \delta \; \alpha \; {\sqrt{2} W \over r} \; f_{4} \; - \;
\delta \; m \; h_{1} = 0
\eqno(2.10c)
$$

\noindent It is easily understandable that one has to
distinguish between two distinct situation: depending on whether
the characteristic  function $W(r)$ is zero or non-zero (we will
remember that the case $W(r)=0$ corresponds to the special simplest
monopole  potential.  The fact  is that the foregoing operator
$\hat{N}_{\alpha}$ with an arbitrary complex-valued $\alpha$ can be
diagonalized upon the functions   (2.9b) if and only if the function
$W(r)$ is equal to zero; otherwise ($W(r) \neq 0$) we have to take
$\alpha = + 1$ (the value $\alpha = -1$ could be chosen as well).

Eventually we obtain the six-equation system (two cases are
distinguished)
$$
 W(r) = 0  \qquad  (\;arbitrary \;\;\alpha \; ) \; ,
$$
$$
\epsilon  f_{3} \; - \; i {d \over dr} \; f_{3}\;
- \; {ib \over r} \; f_{4} \; - \; m\; f_{1} =  0 \; ,
$$
$$
\epsilon
\; f_{4}\; + \; i {d \over dr} \; f_{4}\; + \;({b \over r} \; f_{3}
\; - \;  m \; f_{2}  =  0  \; ,
$$
$$
\epsilon \; f_{1} \; + \; i {d
\over dr} \; f_{1} \; + \; {ib \over r} \; f_{2} \; - \; m \; f_{3} =
0\; ,
$$
$$
\epsilon \; f_{2} \; - \; i {d \over dr} \; f_{2} \; - \;
{ib \over r} \; f_{1} \; - \; m \; f_{4} = 0  \; ,
$$
$$
\epsilon \;
h_{2} \; - \; i {d \over dr} \; h_{2} \; - \; {ia \over r} \; h_{1}
\; - \; \delta \;  m \; h_{1} = 0 \; ,
$$
$$
\epsilon \; h_{1} \; +
\; i {d \over dr} \; h_{1} \; + \; {ia \over r} \; h_{2} \; - \;
\delta \;  m \; h_{2} = 0 \; .
\eqno(2.11a)
$$

\noindent Correspondingly,
$$
W(r) \neq  0  \qquad (\alpha = +1 )
$$
$$
\epsilon \; f_{3} \; - \; i {d \over dr} \; f_{3} \; - \;
{ib \over r} \; f_{4}  \; - \; m \; f_{1} = 0 \; ,
$$
$$
\epsilon \; f_{4} \; + \;  i \; {d \over dr} \; f_{4} \; +
\; {ib \over r}\; f_{3} \; + \;
i \delta \; {\sqrt{2} W \over r} \; h_{2} \; - \; m \; f_{2} = 0 \; ,
$$
$$
\epsilon \; f_{1} \; + \; i {d \over dr} \; f_{1} \; + \;
{ib \over r} \; f_{2}  \; - \; m \; f_{3} = 0   \; ,
$$
$$
\epsilon \; f_{2} \; - \; i {d \over dr} \; f_{2} \; - \;
 {ib \over r} \; f_{1}
\; - \;  i {\sqrt{2} W \over r} \; h_{1}\; - \; m \; f_{4} = 0  \; ,
$$
$$
\epsilon \; h_{2} \; - \; i {d \over dr} \; h_{2} \; - \;
{ia \over r}\; h_{1}   \; - \; i \delta \; {\sqrt{2} W \over r} \;
 f_{4} \;  - \; \delta \; m \; h_{1} = 0 \; ,
$$
$$
\epsilon \; h_{1} \; + \; i {d \over dr} \; h_{1} \; + \;
{ia \over r}\; h_{2} \; + \; i {\sqrt{2} W \over r} \; f_{2} \; - \;
\delta \; m \; h_{2} = 0  \; .
\eqno(2.11b)
$$

It is the point to notice that everything said above is valid as it
stated only  when   $j \ge  3/2$ ; the case of minimal $j = 1/2$
is to be considered separately. First of all, it is a special
substitution for the starting wave function  that should be used:
$$
\Psi ^{0}_{S.}(t,r,\theta ,\phi )\;  = \;
{e^{-i\epsilon t} \over r} \; \times
$$
$$
\left [ \;T_{+1} \otimes \left  ( \begin{array}{c}
 0 \\  f_{2} \; D_{-1/2}  \\  0  \\  f_{4} \; D_{-1/2}
\end{array}  \right ) \;  + \;  T_{0} \otimes
\left ( \begin{array}{c}
h_{1} \; D_{-1/2}  \\ h_{2} \; D_{+1/2}  \\  h_{3} \; D_{-1/2} \\ h_{4} \; D_{+1/2}
\end{array} \right ) \; + \;  T_{-1} \otimes
\left  ( \begin{array}{c}
g_{1} \; D_{+1/2} \\  0  \\  g_{3} \; D_{+1/2} \\  0
\end{array} \right ) \right ] \; .
\eqno(2.12a)
$$

\noindent The angular operator $\Sigma _{\theta ,\phi }$
acts on  $\Psi ^{0} _{S.}(x)$ according to
$$
\Sigma _{\theta ,\phi } \Psi ^{0}_{S.}(x) =
{ e^{-i\epsilon t} \over r}\;
\left [ \; T_{+1} \otimes \left  (  \begin{array}{c} 0   \\  0
\\  0  \\   0  \end{array} \right ) \; +
\;  i \; T_{0} \otimes   \left ( \begin{array}{c}
-h_{4} \; D_{-1/2} \\ +h_{3} \; D_{+1/2} \\
+h_{2} \; D_{-1/2} \\ -h_{1} \; D_{+1/2}
\end{array}        \right ) \;  +  \; T_{-1} \otimes
\left ( \begin{array}{c}
 0 \\ 0 \\  0 \\ 0          \end{array} \right ) \; \right ]
\eqno(2.12b)
$$

\noindent and the corresponding radial system will be
$$
(\epsilon \; + \; \tilde{F} ) \; f_{4} \;  +
i {d \over dr} \; f_{4} \; + \;i{\sqrt{2} W \over r} \; h_{3}\; - \;
(m \; + \; \tilde{\Phi } )\; f_{2} = 0
$$$$
(\epsilon \; + \; \tilde{F})\; f_{2} \; - \;
 i {d \over dr} \; f_{2}\; - \;
i { \sqrt{2} W \over r} \; h_{1} \; - \;
(m \; + \; \tilde{\Phi } )\; f_{4} = 0
\eqno(2.13a)
$$$$
(\epsilon \; - \; \tilde{F}) \; g_{3}\; - \;
i {d \over dr} \; g_{3} \; - \;
i {\sqrt{2} W \over r} \; h_{4} \; - \;
(m \; - \; \tilde{\Phi})\; g_{1} = 0
$$$$
(\epsilon \; - \; \tilde{F})\; g_{1} \; +\;
 i{d \over dr} \; g_{1} \; + \;
i {\sqrt{2} W \over r} \; h_{2} \; - \;
(m \; - \; \tilde{\Phi } )\; g_{3} = 0
\eqno(2.13b)
$$$$
\epsilon \; h_{3} \; - \; i {d \over dr} \; h_{3} \; -\;
{i \over r} \; h_{4}\; - \; i {\sqrt{2} W \over r} \;
 f_{4} \; - \; m \; h_{1} = 0
$$$$
\epsilon \; h_{4} \; + \; i {d \over dr} \; h_{4} \; +
\; {i \over r}\; h_{3}
\; + \;  i {\sqrt{2} W \over r} \; g_{3} \; - \; m\; h_{2} = 0
$$$$
\epsilon \; h_{1} \; + \; i {d\over dr} \; h_{1} \; +
\; {i \over r} \; h_{2}
\; + \; i {\sqrt{2} W \over r} \; f_{2} \; - \; m \; h_{3}  = 0
$$
$$
\epsilon \; h_{2} \; - \; i {d \over dr} \; h_{2} \; - \;
{i \over r} \; h_{1}
\; - \; i {\sqrt{2} W \over r} \; g_{1} \; - \;  m \; h_{4} = 0  \; .
\eqno(2.13c)
$$

\noindent Having taken into account the $\hat{N}_{\alpha}$-based
limitations

$
N = \delta  (-1)^{3/2}  \;\;  , \; \; \delta  = \pm  1 \; :
$
$$
g_{3} = \delta \;  \alpha \;  f_{2} \; , \qquad
g_{1} = \delta \;  \alpha \;  f_{4} \; , \qquad
h_{4} = \delta \; h_{1}   \;  \qquad  h_{3} = \delta \;  h_{2}
\eqno(2.14a)
$$

\noindent we  arrive at  (again $\tilde{\Phi} = 0,
\tilde{F} = 0$ have been set)

$
W(r) = 0 \; , \;\; j = 1/2 \; :
$
$$
\epsilon \;  f_{4} \; + \; i {d \over dr} \; f_{4} \; - \;
m \; f_{2}  =  0 \; ,
$$
$$
\epsilon \; f_{2} \; - \;  i {d \over dr} \; f_{2} \; - \;
m \; f_{4}  =  0\; ,
$$
$$
\epsilon \; h_{2} \; - \; i {d \over dr} \; h_{2} \; -
{i \over r} \; h_{1} \; - \; \delta \;  m \; h_{1} = 0 \; ,
$$
$$
\epsilon  \; h_{1} \; + \; i {d \over dr} \;  h_{1} \; + \;
{i \over r} \; h_{2} \;  - \; \delta \;  m \;  h_{2} = 0    \; .
\eqno(2.14b)
$$

$
W(r) \neq  0 \; , \;\; j = 1/2\; :
$
$$
\epsilon \; f_{4} \; + \;  i {d \over dr} \; f_{4} \; +
\; i \delta \;{\sqrt{2} W \over r} \;  h_{2} \; -
\; m \; f_{2} = 0 \; ,
$$
$$
\epsilon \;  f_{2} \; - \; i {d \over dr} \; f_{2} \; - \;
i {\sqrt{2} W \over r} \; h_{1} \; - \; m \; f_{4} = 0  \; ,
$$
$$
\epsilon \; h_{2} \; - \;  i {d \over dr} \;  h_{2} \;  - \;
{i \over r} \; h_{1}\; - \; i \delta \; {\sqrt{2} W \over r} \;
f_{4} \; - \; \delta  \; m \; h_{1} = 0\; ,
$$
$$
\epsilon \; h_{1} \; + \; i {d \over dr} \;  h_{1} \; + \;
{i \over r} \; h_{2}\;  + \; i   {\sqrt{2} W \over r} \; f_{2}\; -
\; \delta \;  m \; h_{2}  = 0 \; .
\eqno(2.14c)
$$

\subsection*{3.  The operator $\hat{N}_{\alpha}$
                 in  the Dirac and Cartesian gauges}

Now we proceed further with the reflection symmetry and
give some attention to  the question of explicit form of the above
operator $\hat{N}^{S.}_{\alpha}$ (the additional  sign $S.$ of
the Schwinger's isotopic gauge is written) in some other
isotopic gauges. Obviously,  {\em a priori}  there is no  reason to
choose a unique gauge as more preferred that all others; the
particular choice above was  dictated  by convenience only. That
restriction can easily be relaxed, for example, by demonstration of
several more gauges. We are chiefly interested in the Dirac unitary
and Cartesian gauges (as the most frequently used in the literature).

The situation, as a whole, concerning the various representations
being employed in  the present treatment  can be delineated
as follows (here, the designations $Cart.$ and $cycl.$ are
associated with two different bases for a set of isotopic
matrices $t_{i}$; see in Suppl. A)
$$
\Psi ^{C.}_{Cart.} \qquad  \stackrel{\vec{c}}{\Longrightarrow}  \qquad
\Psi ^{D.}_{Cart.} \qquad  \stackrel{\vec{c}\;'}{\Longrightarrow} \qquad
\Psi ^{S.}_{Cart.}  \; ;
$$
$$
\Psi ^{C.}_{cycl.} \qquad \stackrel{V}{\Longrightarrow} \qquad
\Psi ^{D.}_{cycl.} \qquad \stackrel{U}{\Longrightarrow} \qquad
\Psi ^{S.}_{cycl.} \;
$$

\noindent those functions are connected by the relations
$$
\Psi ^{D.}_{Cart.} = O(\vec{c})\; \Psi ^{C.}_{Cart.} \; ,\qquad
\Psi ^{S.}_{Cart.} = O(\vec{c}\;') \; \Psi ^{D.}_{Cart.}\;  ,
$$$$
\Psi ^{C.}_{cycl.} = S \; \Psi^{C.}_{Cart.}  \; , \qquad
\Psi ^{D.}_{cycl.} = S \; \Psi ^{D}_{Cart.}  \; , \qquad
\Psi ^{S.}_{cycl.} = S \; \Psi ^{S.}_{Cart.}
\eqno(3.1)
$$

\noindent where $\vec{c}$  is the known Gibbs parameter on the group
$SO(3.R)$ [64])
$$
O(\vec{c})  = \left ( \begin{array}{ccc}
[1 - (1 - \cos \theta )\cos^{2} \phi] & -(1 -\cos \theta) \sin\theta
\cos \theta  &  - \sin \theta  \cos \phi  \\
-(1  - \cos \theta ) \sin \theta  \cos \phi  &
[1  - (1 - \cos \theta ) \sin ^{2} \phi ] &  - \sin \theta  \sin \phi  \\
\sin \theta  \cos \phi  & \sin \theta  \sin \phi  & \cos \theta
\end{array} \right  )
$$
$$
O(\vec{c}\; ') = \left ( \begin{array}{ccc}
\cos \phi  & \sin \phi  & 0 \\
-\sin \phi  & \cos \phi  & 0 \\  0  & 0 & 1
\end{array} \right ) \; , \qquad
S = \left ( \begin{array}{ccc}
-1/ \sqrt{2} & i/ \sqrt{2} & 0   \\    0 & 0 & 1  \\
+1/ \sqrt{2} &  i/ \sqrt{2}  & 0
\end{array} \right )
$$

\noindent  Above in the work, the $S.$-gauge (Schwinger's) and
{\em cyclic} basis  solely were  used. The
passage to the Dirac gauge can be with no difficulty performed
$$
\Psi ^{D.}_{cycl.} = U \; \Psi ^{S.}_{cycl.} \; , \qquad
U = S \; O(-\vec{c}\; ') \; S^{-1}
\eqno(3.2a)
$$

\noindent for the matrix $U$ we get
$$
U(\phi)  = \left ( \begin{array}{ccc}
e^{-i\phi }  &  0  &  0  \\  0  &  1  &   0   \\  0    &  0  & e^{+i\phi}
\end{array} \right )  \; .
\eqno(3.2b)
$$

\noindent    Correspondingly, after translating to the $D.$-gauge
$  \hat{N}^{D.}_{cycl.} = U(\phi ) \; \hat{N}^{S.}_{cycl.} \;  U^{-1}(\phi ) \; $,
the  discrete operator  $\hat{N}_{\alpha }$  takes the form
$$
\hat{N}^{D.}_{cycl.} = [\; ( U^{-1} \; \hat{\pi }^{S.}_{\alpha } \; U ) \;
U_{0} \otimes  ( - \gamma ^{5} \gamma ^{1} )\; ] \otimes \hat{P}
\eqno(3.2c)
$$

\noindent where  $U_{0}$ is a matrix arising out of the commutation rule
$\hat{P} \; U(\phi ) = U(\phi ) \; U_{0} \; \hat{P}$ and given by
$$
U_{0} = \left ( \begin{array}{rrr}
-1 & 0 &  0 \\ 0  & +1  &  0 \\ 0 &  0 & -1
\end{array}  \right )              \; .
$$

\noindent
Ultimately, for  $\hat{N}^{D.}_{cycl.}$ we find the following
explicit form
$$
\hat{N}^{D.}_{cycl.} = ( \hat{\pi }^{D.}_{cycl.} \otimes
(- \gamma ^{5} \gamma ^{1}) ) \otimes \hat{P}
$$

\noindent  where the matrix  $\hat{\pi }^{D.}_{cycl.}$ is given by
$$
\hat{\pi }^{D.}_{cycl.} =
\left ( \begin{array}{ccc}
 0 & 0 & (\alpha ^{-1}) \; e^{-2i\phi } \\
0 & 1  &  0  \\    \alpha  \; e^{+2i\phi } & 0 & 0
\end{array} \right )   \; .
\eqno(3.2d)
$$

In the same way we find the determining relations for the Cartesian gauge
$$
\Psi ^{Cart.}_{cycl.} = U(\theta ,\phi ) \; \Psi ^{S.}_{cycl.}\; , \;\;
\eqno(3.3a)
$$
$$
U(\theta ,\phi ) = \left ( \begin{array}{ccc}
{1 \over 2} (1 + \cos \theta ) e^{-i\phi } &  { 1 \over \sqrt{2}} \sin \theta
e^{-i\phi } & {1 \over 2} (1-\cos  \theta ) e^{-i\phi } \\
{1 \over \sqrt{2}} \sin \theta  & \cos \theta &
-{1 \over \sqrt{2}} \sin \theta  \\
 (1-\cos  \theta ) e^{+i\phi } & {1 \over \sqrt{2}} \sin \theta  e^{+i\phi } &
{1 \over \sqrt{2}} (1-\cos \theta ) e^{-i\phi }
\end{array} \right )   \; .
$$

\noindent Decomposing the operator  $\hat{\pi }^{S.\alpha }_{cycl.}$ into
a product of two factors ($\alpha  \equiv e^{iA}$, $A$ is a complex number)
$$
\hat{\pi }^{S.\alpha }_{cycl. } =
\left ( \begin{array}{ccc}
0 & 0 & \alpha ^{-1} \\
0 & 1 & 0  \\  \alpha  & 0 & 0
\end{array} \right ) = \left ( \begin{array}{ccc}
e^{-iA} & 0 & 0  \\  0 & 1 & 0   \\   0 & 0 & e^{+iA}
\end{array} \right ) \;
\left ( \begin{array}{ccc}  0 & 0 & 1 \\ 0 & 1 & 0 \\ 1 & 0 &
0 \end{array} \right ) \equiv
\Delta (A) \;  \hat{\pi }^{S.}_{cycl.}
$$

\noindent  for the $N$-operator in Cartesian gauge
we get the factorization
$$
\hat{\pi }^{Cart.A} _{cycl.} = \left (\; U(\theta ,\phi ) \; \Delta (A) \;
U^{-1}(\theta ,\phi )\; \right )\;
\left (\; U(\theta ,\phi ) \; \hat{\pi }^{S.}_{cycl.} \;
U^{-1}(\pi - \theta ,\phi  + \pi )\; \right )  \; .
\eqno(3.3b)
$$

\noindent The second term-multiplier can be brought to the form
$$
  U(\theta ,\phi ) \; \hat{\pi }^{S.}_{cycl.} \;
U^{-1} (\pi-\theta ,\phi + \pi )  = - I \;
\eqno(3.4)
$$

\noindent and the first one may be rewritten as
$$
U(\theta ,\phi ) \; \Delta (A) \; U^{-1}(\theta ,\phi ) =
\left ( S \; O(- \vec{c}\; ') \; S^{-1} \right ) \;
\Delta (A) \;
\left ( S \;  O( \vec{c} \; ' ) \; S^{-1} \right ) \; .
$$

\noindent Now, noting    the identity
$$
S^{-1} \; \Delta (A) \; S = \left ( \begin{array}{ccc}
\cos A & - \sin A & 0 \\  - \sin A & \cos A & 0 \\ 0 & 0 & 1
\end{array} \right ) \equiv  O(\vec{a}) \; ,
\qquad \vec{a} = ( 0, 0, -\tan A/2)
 $$

\noindent where $(O(\vec{a})$  is    a transformation of the complex rotation
group, and also taking into account the known in the
theory of the group $SO(3.$') [64]  relation
$$
O(-\vec{c}\; '') \; O( \vec{a}) \; O(\vec{c}\; '') =
\; O ( O_{-\vec{c}\; ''} \vec{a} )
$$

\noindent one can get
$$
U(\theta ,\phi ) \; \Delta (A) \; U^{-1}(\theta ,\phi ) =
S \; O ( O_{-\vec{c}\; ''} \vec{a}) \; S^{-1} \; .
\eqno(3.5a)
$$

\noindent The vector  $(O_{\vec{c}\; ''} \vec{a})$   has the
following explicit form
$$
( 0_{-\vec{c}\; ''} \vec{a} ) = - \tan A/2 \; \left (\; \sin \theta  \cos \phi ,
\; \sin \theta  \sin \phi, \; \cos \theta \; \right )
\eqno(3.5b)
$$

\noindent so the expression for
$ S\; O( 0_{-\vec{c}''} \vec{a}) \; S^{-1}$  is as follows
$$
S \; 0 (0_{-\vec{c}\; ''} \vec{a}) \;  S^{-1} =
S \;  O(-\tan A/2 \; \vec{n}_{\theta ,\phi }) \;  S^{-1} =
$$
$$
S \;\; \exp  \left [\; i \;A \; \vec{j} \; \vec{n} _{\theta ,\phi }\;
 \right ] \;\;  S^{-1} =
\exp \left [\; i \; A \; \vec{t}\; \vec{n} _{\theta ,\phi }\; \right ] \;
\eqno(3.5c)
$$

\noindent
here we employ the well-known exponential representation  of orthogonal
$3\times3$-matrix (for more details see [64]).
Given these relationships  (3.4) ¨ (3.5c), one can find
$\hat{N}_{A}$-operator in Cartesian gauge
$$
\hat{N}^{Cart.A}_{cycl.} =   \hat{\pi }^{Cart.A} _{cycl.}  \otimes
 ( -\gamma ^{5} \gamma ^{1}) )  \otimes  \hat{P} =
= \exp \left ( \;i\; A\; \vec{t} \; \vec{n}_{\theta ,\phi} \; \right )
 \otimes    (- \gamma ^{5} \gamma ^{1})  \otimes \hat{P} \; .
\eqno(3.6a)
$$

\noindent In the case  $\alpha  = 1$, from  (3.6a) it follows
$$
\alpha = + 1 \; , \qquad
\hat{\pi }^{Cart.}_{cycl.} = - I   \; .
\eqno(3.6b)
$$

\noindent
The  accomplished expression for  $\hat{\pi }^{Cart.}_{cycl.}$
shows that the ordinary $P$-reflection operator for a single bispinor field
can be diagonalized on the composite (triplet's) wave functions (when $A\equiv 0$).
This may also  be demonstrated by direct calculation over the explicit wave functions
$\Psi ^{Cart.}_{cycl.}$ in this gauge.
The explicit expression for $\Psi ^{Cart.}_{cycl.}$ is
$$
\Psi ^{Cart.}_{cycl.} = { e^{-i\epsilon t} \over r}  \; \times
$$
$$
\left \{  T_{+1} \otimes  e^{-i\phi } \;
\left [ \; {1 + \cos \theta  \over 2 } \; F \; - \;
                {\sin \theta  \over 2} \; H \; + \;
           {1 - \cos  \theta  \over 2} \; G \; \right ]  \; + \right.
$$
$$
+ \; T_{0} \otimes \left [\; { \sin \theta  \over \sqrt{2} }  \; F
\; + \; \cos \theta  \; H \; - \;
{\sin \theta  \over\sqrt{2}} \; G \; \right ]\; +
$$
$$
+ \; T_{-1} \otimes e^{+i\phi } \left. \left [ \; {1 - \cos \theta
\over 2 } \; F \; + \; { \sin \theta  \over 2} \; H \; + \;
 {1 + \cos  \theta  \over 2} \; G \; \right ] \right \}
\eqno(3.7a)
$$

\noindent here the symbols  $(F,\; H, \;G)$  denote
4-component column-functions
$$
F = \left ( \begin{array}{c}
f_{1} \; D_{-3/2} \\ f_{2} \; D_{-1/2} \\
f_{3} \; D_{-3/2} \\ f_{4} \; D_{-1/2}   \end{array} \right ) \; , \;\;
H = \left ( \begin{array}{c}
h_{1} \; D_{-1/2} \\ h_{2} \; D_{+1/2} \\
h_{3} \; D_{-1/2} \\ h_{4} \; D_{+1/2}
\end{array} \right ) \; , \;\;
G = \left ( \begin{array}{c}
g_{1} \; D_{+1/2}  \\  g_{2} \; D_{+3/2}  \\
g_{3} \; D_{+1/2}  \\  g_{4} \; D_{+3/2}
\end{array}  \right  )  \; .
\eqno(3.7b)
$$

\noindent After simple calculation for the proper values equation
$$
[\; I \otimes  (-\gamma ^{5} \; \gamma ^{1}) \otimes  \hat{P} \; ]
\; \Psi ^{Cart.}_{cycl.} \; = \;
P \; \Psi ^{Cart.}_{cycl.}
\eqno(3.8)
$$

\noindent  one can arrive at the same relations  (2.9a).

\begin{quotation}

{\it
It is  important to point out that from the diagonalization  of this
discrete operator one cannot  infer that the whole problem of reflection symmetry of
the non-Abelian monopole amounts    to the Abelian one.
Indeed, the non-Abelian $N$-operator acts on  the functions
(3.7b), but  the  function $F(x),H(x),G(x)$ carrying individual Abelian
properties are just represented in composites functions as some constructing elements,
so that the function-multipliers at $T_{-1},T_{0},T_{+1}$ cannot be induced,
via any special $U(1)$-gauge transformation, from Abelian ones.
The operation of changing the isotopic gauge  $S. \Longrightarrow Cart.$
may be regarded as the calculated modification of isotopic gauge in order to
translate  a non-null transformation upon the isotopic coordinates
in Schwinger gauge into  the null transformation upon these coordinates
in the Cartesian gauge. But a natural consequence of this is that the all individual
Abelian qualities of the   $T_{-1},T_{0},T_{+1}$'s function-multipliers
have vanished completely. }

\end{quotation}

\subsection*{4. $\hat{N}_{A}$-parity selection rules.}

Now we proceed to analyzing  the question of what is the meaning of the parameter
$A$. Any value of this $A$  specifies one of the possible operators of
the complete   set of variables; besides, the corresponding mark appears in the
designation and expression of the relevant wave functions
$\Psi ^{\alpha }_{\epsilon jm\delta }$
and  $\Psi ^{\alpha }_{\epsilon 0\delta }$ (below the quantum numbers
$\epsilon ,j,m$   are omitted; for definiteness supposing  $j \ge  3/2$)
$$
\Psi ^{S.A }_{\delta \mu } (x) =
\left [ \;  T_{0} \otimes \Phi _{\delta }(x) \; + \;
T_{+1} \otimes \Phi ^{+}_{\mu}(x) \; + \; \delta \;\mu \;  e^{iA} \;
T_{-1} \otimes \Phi ^{-}_{\mu} (x) \;  \right ] \;.
\eqno(4.1)
$$

\noindent
The complex parameter  $A$  gives  a limitation on freedom with  which
the one triplet component proportional  to $T_{-1}$ is built from another
proportional  to  $T_{+1}$. Every $N_{A}$-operator  with a fixed $A$ prescribes
exactly how three isotopic  components  link up  with each other, so that a
unique whole is produced  via a kinematical  association of dynamically
independent subsystems. This characteristic feature of the triplet wave  function
 at $W(r) = 0$, as evidenced from the previous treatment,  is formalized
through  just the $N_{A}$-operator. It is noticeable that such a freedom
refers solely to the couple of  $T_{-1}$  and $T_{+1}$.

The presence of the $A$-parameter in the explicit expressions of the
wave function  finds its  corollary in possible manifestation over
matrix elements of physical quantities.  Let $\hat{G}$   be a certain
composite observable, and the triplet functions are written schematically
as
$$
\Psi ^{\alpha }_{\delta \mu }  =  (\; \Psi ^{0}_{\delta} \; + \;
\Psi ^{+}_{\mu} \; + \; \delta \;\mu \; e^{iA} \; \Psi ^{-}_{\mu}\; )  \;
\eqno(4.2a)
$$

\noindent
then for the $\hat{G}$'s expectation value we can easily obtain the  following
expansion (the requirement of Hermiticity is imposed on $\hat{G}$)
$$
G \; = \; < \Psi ^{A}_{\delta \mu } \mid  \hat{G} \mid
\Psi ^{A}_{\delta \mu } > \; = \; < \Psi ^{+}_{\mu} \mid \hat{G}\mid
\Psi ^{+}_{\mu} > \;  +
$$
$$
<\Psi ^{0}_{\delta} \mid \hat{G}\mid  \Psi ^{0}_{\delta} > \;  +
\; 2\;  Re \; <\Psi ^{+}_{\mu} \mid \hat{G}\mid
\Psi ^{0}_{\delta} > \; + \;  e^{i(A-A^{*})} \;
 <\Psi ^{-}_{\mu} \mid \hat{G} \mid  \Psi ^{-} _{\mu}> \; +   \;
$$
$$
2\; \delta \; \mu \; Re \; \left (\; e^{iA} \;
<\Psi ^{+}_{\mu} \mid \hat{G} \mid \Psi ^{-} _{\mu} > \; \right ) \;  +  \;
2 \delta \; \mu \; Re \; \left (\; e^{iA} \;
< \Psi ^{0}_{\delta} \mid \hat{G}\mid  \Psi ^{-}_{\mu} > \; \right ) \; .
\eqno(4.2b)
$$

As a simple  illustration  let us consider in more detail the problem of
$\hat{N}_{A}$-parity selection rules,   particularly giving attention to
their $A$-dependence. An initial matrix element may be written in the form
$$
\int \bar{\Psi}^{A}_{\epsilon JM\delta }(\vec{x}) \; \hat{G}(\vec{x}) \;
     \Phi ^{A}_{\epsilon J'M'\delta '}(\vec{x})\; dV\; \equiv \;
\int r^{2}\; dr \int f^{A}(\vec{x})\; d \Omega \;  .
$$

\noindent
Taking into account the relation
$$
f^{A}(-\vec{x}) = \delta \; \delta' \; (-1)^{J+J'}   \;
\bar{\Psi}^{A}_{\epsilon JM\delta }(\vec{x})  \; \left [ \;
( \hat{N}^{+}_{A} \otimes  \hat{}) \; \hat{G}(-\vec{x}) \;
( \hat{N}_{A} \otimes \hat{}) \; \right ] \;
\Psi ^{A}_{\epsilon J'M'\delta} (\vec{x}) \;
\eqno(4.3a)
$$

\noindent and supposing that $\hat{G}(\vec{x})$ satisfies
$$
\left [ \; \left ( \hat{N}^{+}_{A} \otimes \hat{} \right )
\;  \hat{G}(-\vec{x}) \;  \left ( \; \hat{N}_{A} \otimes \hat{}
\right ) \right ]  = \Omega ^{A} \;  \hat{G}(\vec{x})
\eqno(4.3b)
$$

\noindent where $\Omega ^{A} = + 1$ or $ -1$, we get
$$
f^{A}(-\vec{x}) = \Omega ^{A}\; \delta  \; \delta' \;
(-1)^{J+J'}\; f^{A}(\vec{x}) \; ,
\eqno(4.3c)
$$

\noindent  Therefore, the integral from (4.3a) can be transformed into
$$
\int \bar{\Psi}^{A}_{\epsilon JM\delta }(\vec{x}) \; \hat{G}(\vec{x}) \;
     \Phi ^{A}_{\epsilon J'M'\delta '}(\vec{x})\; dV\; =
$$
$$
= \left [ \; 1\;  + \; \Omega^{A} \; \delta \; \delta ' \; (-1)^{J+J'} \; \right ]
\; \int_{V_{1/2}}
\bar{\Psi}^{A}_{\epsilon JM\delta }(\vec{x}) \; \hat{G}(\vec{x}) \;
\Phi ^{A}_{\epsilon J'M'\delta '}(\vec{x})\; dV\; \equiv \;
\eqno(4.3d)
$$

\noindent
where the symbol $V_{1/2}$ denotes the half-space of integration. The expansion
(4.3c) furnishes, in fact, the required selection rules: if $\delta, \delta', J, J'$
are chosen such that  factor
$( \; 1\;  + \; \Omega^{A} \; \delta \; \delta ' \; (-1)^{J+J'} \;)$ is equal to zero
then the corresponding matrix element  is null too.

The above condition (4.3b) is, in fact, a definition of composite scalars and
pseudoscalars under the operation of $N_{A}$-inversion.  The (4.3b)  can
be easily enlarged on. To this end, let us introduce more detailed notation
for $\hat{G}$:
$$
\hat{G}(\vec{x}) =   \left ( \begin{array}{ccc}
\hat{g}_{11}(\vec{x})  & \hat{g}_{12}(\vec{x})  &  \hat{g}_{13}(\vec{x})  \\
\hat{g}_{21}(\vec{x})  & \hat{g}_{22}(\vec{x})  &  \hat{g}_{23}(\vec{x})  \\
\hat{g}_{31}(\vec{x})  & \hat{g}_{32}(\vec{x})  &  \hat{g}_{33}(\vec{x})
\end{array}   \right  )  \otimes \hat{G}_{0}(\vec{x})
\eqno(4.4a)
$$

\noindent
then the condition  (4.3b) takes the form
$$
 \left ( \begin{array}{rrr}
e^{i(A-A^{*})} \; \hat{g}_{33}(-\vec{x}) &
e^{-iA^{*}} \; \hat{g}_{32}(\vec{x})     &
e^{-i(A+A^{*})} \hat{g}_{31}(-\vec{x})    \\
e^{+iA} \; \hat{g}_{23}(-\vec{x})        &
\hat{g}_{22}(-\vec{x})  & e^{-iA} \;\hat{g}_{21}(-\vec{x}) \\
e^{i(A+A^{*})} \; \hat{g}_{13}(-\vec{x})  &
e^{+iA^{*}} \; \hat{g}_{12}(- \vec{x})   &
e^{-i(A-A^{*})} \; \hat{g}_{11}(-\vec{x})
\end{array} \right ) \otimes
$$
$$
\otimes \;
\hat{} \; \hat{G}_{0}(-\vec{x}) \; \hat{} = \; \Omega ^{A} \; \left
( \begin{array}{ccc}
\hat{g}_{11}(\vec{x}) & \hat{g}_{12}(\vec{x}) & \hat{g}_{13}(\vec{x}) \\
\hat{g}_{21}(\vec{x}) & \hat{g}_{22}(\vec{x}) & \hat{g}_{23}(\vec{x}) \\
\hat{g}_{31}(\vec{x}) & \hat{g}_{32}(\vec{x}) & \hat{g}_{33}(\vec{x})
\end{array} \right )    \; \hat{G}_{0}(\vec{x})   \; .
\eqno(4.4b)
 $$

\subsection*{5. On relating the $\Psi ^{A}_{\epsilon jm\delta }$
at different $A$'s}

All different values of $A$ give the same Hilbert space of quantum states.
The transformation from $A$- to $A'$-basis is described in the Schwinger's gauge
as follows:
$$
\Psi ^{A'}_{\epsilon jm\delta } = \left [ \;
e^{-iA} \left ( \begin{array}{ccc}
e^{iA} & 0 & 0 \\
0 & e^{iA} & 0 \\
0 & 0 & e^{iA'}
\end{array} \right ) \otimes I \right ] \; \Psi ^{A}_{\epsilon jm\delta } \;
\eqno(5.1a)
$$

\noindent   (further we follow only its isotopic part).
Let us look into  a general structure of the relationship
$$
\Psi ^{A'}_{S.}(x) = V_{S.} (A',A)\; \Psi ^{A} _{S.}(x)\; .
\eqno(5.1b)
$$

\noindent  To this end, factorizing  the $ V(A',A)$    according to
( $ \Gamma \equiv (A' - A) / 2 $ )
$$
V_{S.}(A',A) \equiv  e^{+i\Gamma } \left ( \begin{array}{ccc}
1   & 0             & 0  \\
0   & e^{-i\Gamma } & 0  \\
0   &  0            & 1
\end{array} \right ) \left ( \begin{array}{ccc}
e^{-i\Gamma }& 0 & 0 \\
0 & 1 & 0 \\
0 & 0 & e^{+i\Gamma }
\end{array} \right )
\eqno(5.2a)
$$

\noindent or
$$
V_{S.}(A', A)  \equiv   e^{+i\Gamma } \; D(\Gamma ) \; \Delta (\Gamma ) \; .
\eqno(5.2b)
$$

\noindent The term  $\Delta (\Gamma )$ already appeared; it  represents a rotation
of angle $\Gamma$ about  third isotopic axis $(0,0,1)$.
The second transformation  $D(\Gamma )$ acts only on the vector
$T_{0}$  in isotopic space, and it changes its length trough multiplying
it by a complex number (it seems to be referred to the so-called conformal
ones). The matrix  $D(\Gamma )$  may be rewritten in the exponential form
$$
D(\Gamma ) = e^{-i\Gamma t^{0}}, \qquad  t^{0} =
\left ( \begin{array}{ccc}
0 & 0 & 0 \\ 0 & 1 & 0 \\ 0 & 0 & 0 \end{array} \right ) \; .
\eqno(5.3 )
$$

\noindent  With the use of  explicit expressions for  the isotopic rotation
generators $t_{i}$ and their squares
$$ t_{1} =  {1 \over \sqrt{2}} \left ( \begin{array}{ccc}
0 & 1 & 0 \\ 1 & 0 & 1 \\ 0 & 1 & 0 \end{array} \right ) \; ,  \;
t_{2} = {1 \over \sqrt{2}} \left ( \begin{array}{ccc}
0 & -i & 0 \\ i & 0 & -i \\ 0 & i & 0 \end{array} \right ) \; , \;
t_{3} =  {1 \over \sqrt{2}} \left ( \begin{array}{ccc}
1 & 0 & 0 \\ 0 & 0 & 0 \\ 0 & 0 & -1 \end{array} \right ) \; ,
$$
$$
( t_{1})^{2} = \left ( \begin{array}{ccc}
{1\over 2} & 0 & {1 \over 2} \\ 0 & 1 & 0 \\ {1 \over 2} & 0 & {1\over 2}
 \end{array} \right ) \; , \;
( t_{2})^{2} = \left ( \begin{array}{ccc}
{1 \over 2} & 0 & -{1 \over 2} \\ 0 & 1 & 0 \\
-{1\over 2} & 0 & {1 \over 2} \end{array} \right ) \; , \;
( t_{3})^{2} = \left ( \begin{array}{ccc}
1 & 0 & 0 \\ 0 & 0 & 0 \\ 0 & 0 & 1 \end{array} \right )
$$

\noindent we can produce a formula for $t^{0}$
$$
t^{0} = {1 \over 2} (\; t^{2}_{1} \; + \;  t^{2}_{2} \; - \; t^{2}_{3}\; ) .
\eqno(5.3b)
$$

\noindent Now let us find a representation for  $V(A', A)$ in the Cartesian gauge
$$
V_{Cart.}(A', A) \equiv  e^{+i\Gamma } \;
\left [ \; U(\theta ,\phi ) \; D(\Gamma ) \; U^{-1}(\theta ,\phi )\;\right ]
\; \left [ \;  U(\theta ,\phi ) \; \Delta (\Gamma ) \;
U^{-1}(\theta ,\phi )\; \right ] \; .
\eqno(5.4a)
$$

\noindent The second term in brackets has been already  calculated
$$
U(\theta ,\phi ) \; \Delta (\Gamma ) \; U^{-1}(\theta ,\phi ) \; = \;
\exp  [\;i \; \Gamma \; \vec{t}\; \vec{n}_{\theta ,\phi }\;] \;  .
\eqno(5.4b)
$$

\noindent At analyzing  the third tem it is convenient to utilize
the exponential representation for $D(\Gamma )$ and also the identity
(well-known in the theory of 3-rotation [64])
$$
U(\vec{n} ) \; t_{i} \; U(-\vec{n}) \; = \;
\; U_{ij}(\vec{n}) \; t_{j}\;  \equiv \; \tilde{t}_{i}(\theta ,\phi ) \; .
\eqno(5.4c)
$$

\noindent Ultimately, the $V_{Cart.}(A', A)$  is written as
$$
V_{Cart.}(A', A) \;  \equiv  \; e^{+i\Gamma }  \;
\exp  \left [ \; -i\; \Gamma \; \tilde{t} ^{0}(\theta ,\phi ) \; \right ] \;
\exp  \left [ \;i \;\Gamma \; \vec{t} \; \vec{n} _{\theta ,\phi }\; \right ] \; .
\eqno(5.4d)
$$

\noindent Here, the first term has the trivial $U(1)$ character,
the second  is a conformal one; the third  describes a rotation
of complex 3-dimensional group $S0(3,C)$ about the axis
$\vec{n} _{\theta ,\phi }$.

Let us  detail an explicit expression for  this conformal transformation
in Cartesian  gauge.  Taking into account
that the relation $ (\tilde{t}^{0}) ^{2} = \tilde{t}^{0}$, we can obtain
the following expansion
$$
D^{Cart.}_{cycl.}(\Gamma ) =
U(\theta ,\phi ) \; D(\Gamma ) \; U^{-1}(\theta ,\phi ) \; = \;
[\; I \; + \; ( e^{-i\Gamma } - 1 ) \;\tilde{t}^{0}\; ] \;
\eqno(5.5)
$$

\noindent and, on straightforward calculation, for the
$\tilde{t}^{0}$ we get
$$
\tilde{t}^{0} = \left (  \begin{array}{ccc}
{1 \over 2} \sin^{2} \theta   &
- {1 \over \sqrt{2}} \sin \theta  \cos \theta  e^{-i\phi }   &
-{1 \over 2} \sin ^{2}\theta  e^{-2i\phi }                   \\
-{1 \over \sqrt{2}} \sin \theta  \cos \theta  e^{+i\phi }    &
\cos ^{2}\theta                                              &
{1 \over \sqrt{2}} \sin \theta  \cos \theta  e^{-i\phi }     \\
-{1 \over 2} \sin ^{2}\theta  e^{+2i\phi }                   &
{1 \over \sqrt{2}} \sin \theta  \cos \theta  e^{+i\phi }     &
{1 \over 2}   \sin ^{2}\theta
\end{array} \right )    \; .
$$

\noindent As a controlling test on  $\tilde{t}^{0}$, we can verify the
identity  $( \tilde{t}^{0})^{2} = \tilde{t}^{0}$.

Now  let us  turn again to  the structure of the transformation
$V_{S.}(A', A)$ :
$$
\Psi ^{A'}_{S.}(x) = V_{S.}(A', A) \;\Psi ^{A}_{S.}(x) \; , \qquad
V_{S.}(A', A)  \equiv   e^{+i\Gamma } \; D(\Gamma ) \; \Delta (\Gamma ) \;
\eqno(5.6a)
$$

\noindent remembering $\Gamma = (A' - A)/2$ .
In accordance with the definition, the relationship
$$
V_{S.} (A', A) \; \hat{N}^{A}_{S.} \; V^{-1}_{S.}(A', A) =
\hat{N}^{A'}_{S.}\;  .
\eqno(5.6b)
$$

\noindent must hold. However,  allowing for the above factorization  of
$V_{S.} (A',A)$, it is readily verified that only the operator
$\Delta (\Gamma )$ ( a single term from two ones)
makes actually this translation:
$\hat{N}^{A}_{S.} \Longrightarrow \hat{N}^{A'}_{S.} $,
whereas the remaining term  do not:
$D(\Gamma ) \;\hat{N}^{A}_{S.} \; D^{-1}(\Gamma ) =
\hat{N}^{A}_{S.}$. One should remember
that   $D(\Gamma )$  acts only on  the isotopic   component proportional to
$T_{0}$.  Moreover, it is evident that $D(\Gamma )$ does not affect all the  operators
from  a complete set of variables: $ i\partial_{t}, \vec{J}^{\;2}, J_{3}$
as well as the $\hat{N}_{A}$.

The existence of an operation with  those properties indicates that  there exists a
possibility  to subject the function
 $\Psi ^{A}_{\epsilon jm\delta }(x)$ to a  transformation of  this
$D(F)$ kind (here $F$ involved may not  be correlated
with  the above  $\Gamma$), so that  as a result  we arrive at a new wave function
 with the same quantum numbers. However this situation does not provide us with
a paradox. To understand this,  it suffices to return  to the radial system
(2.11a) and give attention to the fact that in this  system the two sub-sets
of functions $f_{1}, \; f_{2}, \; f_{3}, \; f_{4}$  and
$h_{1}, \; h_{2}$ are completely independent of each other.
The latter speaks  of the following:
the set of operators used above and provided the quantum number
$(\epsilon, j,m,\delta)$ fixes the wave function apart from an arbitrary
complex factor at the multiplet component proportional to $T_{0}$
$$
\Psi ^{A(B)}_{\epsilon jm\delta }( \vec{x}) =  \left [ \;
T_{+1} \otimes \Phi ^{+}(\vec{x}) \; + \;
e^{iB} \; T_{0} \otimes \Phi _{\delta }(\vec{x}) \; + \;
\delta  \; e^{iA} \; T_{-1} \otimes \Phi ^{-}(\vec{x}) \; \right ]
\eqno(5.7a)
$$

\noindent The  transformation of $D$-kind acts on those functions according to
$$
\Psi _{\epsilon jm\delta \mu }^{A(B')}(\vec{x}) = D(B - B')\;
\Psi ^{A(B)}_{\epsilon jm\delta \mu }(\vec{x}) \; .
\eqno(5.7b)
$$

\noindent    Evidently, such a $B$-symmetry can prove itself in explicit
expression   for matrix elements; but this  cannot happen to the
$N$-parity selection rules because this $D$-symmetry does not affect the
$N$-operator.

\newpage

\subsection*{ Supplement A . Some technical material:
Dirac and Schwinger gauges in isotopic space; spinor approach by
Tetrode-Weyl-Fock-Ivanenko;  Schr\"{o}dinger's basis and Pauli's analysis.}

This  section deals with some representation of the
non-Abelian monopole potential,  which  will  be  the  most
convenient one to formulate and analyze the problem of isotopic
multiplet in this field.
Let us begin describing in detail this matter. The well-known
form  of the monopole solution  introduced  by t'Hooft and Polyakov
([61,62]; see also Julia-Zee [63]) may be taken as a starting point.
The field   $W^{(a)}_{\alpha }$   represents a~covariant vector with
the~usual transformation law
$W^{(a)}_{\beta } = (\partial x^{i} / \partial x^{\beta }) W^{(a)}_{i})$
and our first step  is the~change of variables in 3-space.
Thus, the~given  potentials
$( \Phi ^{(a)}(x), W^{(a)}_{\alpha })$ convert into
$ (\Phi ^{(a)}(x), W^{(a)}_{t}, W^{(a)}_{r}$,
$W^{(a)}_{\theta }, W^{(a)}_{\phi})$.
Our second step is a~special  gauge  transformation  in  the
isotopic space. The~required gauge matrix can be determined  (only
partly) by the~condition
$\;(O_{ab} \Phi ^{b}(x) ) = ( 0 , 0 , r \Phi (r)\; )$.
This equation has a~set of solutions  since the~isotopic
rotation by every angle about the~third axis   $(0, 0, 1)$   will  not
change the~finishing vector $( 0, 0, r \Phi (r) )$. We fix
such  an~ambiguity  by  deciding  in  favor  of   the~simplest
transformation matrix. It will be convenient to utilize  the~known
group $SO(3.R)$ parameterization through the Gibbs
3-vector\footnote{The author highly
recommends the book [64] for many further details
developing  the Gibbs approach to groups
$SO(3.R), SO(3.C), SO_{0}(3.1)$, etc.}
$$
O = O( \vec{c} ) =  I + 2 \;{ \vec{c}^{\; \times } + ( \vec{c}^{\; \times })^{2} \over
 1 + \vec{c}^{2}} \;\; ,  \qquad
 (\vec{c}^{\; \times })_{ac} = - \epsilon _{acb} \; c_{b} \; .
\eqno(A.1)
$$

\noindent According to [65], the simplest rotation above is
$$
\vec{B} = O(\vec{c}) \vec{A} \;  , \; \;
\vec{c} =  { [\vec{B} \vec{A} ] \over   (\vec{A} +\vec{B} ) \vec{A}} \; .
$$

\noindent Therefore,
$$
\hbox{if}\;\;\;\; \vec{A} = r \Phi (r) \;
\vec{n}_{\theta, \phi} \;  , \; \;
\vec{B} = r \Phi (r) ( 0 , 0 , 1 )\; , \; \;
$$
$$
\hbox{then}  \qquad
\vec{c} = {{\sin \theta }\over {1 + \cos \theta }} \;
( + \sin \phi  , - \cos \phi  , 0 )    \; .
\eqno(A.2)
$$

\noindent Together with  varying the~scalar  field $\Phi ^{a}(x)$, the~vector
triplet $W^{(a)}_{\beta }(x)$  is to be transformed from one isotopic
gauge  to another under the~law  [66]
$$
W'^{(a)}_{\alpha }(x) \; = \; O_{ab} (\vec{c}(x)) \; W^{(b)}_{\alpha }(x) \; + \;
{1\over e} \; f_{ab}( \vec{c}(x))\; {{\partial c_{b} } \over
 { \partial x^{\alpha }}} \;\; ,  \qquad
f(\vec{c}) = - 2\; { 1 + \vec{c}^{\; \times }  \over
1 + \vec{c}^{\; 2}} \;\; .
\eqno(A.3)
$$

\noindent With  the  use  of (A.3), we obtain  the new representation
$$
\Phi^{D.(a)}= r \Phi (r)
\left ( \begin{array}{c}
   0 \\ 0 \\ 1
\end{array} \right )\; , \qquad
              W^{D.(a)}_{\theta } = (r^{2} K + 1/e)
\left ( \begin{array}{c}
              - \sin \phi \\
              + \cos \phi  \\
                0
\end{array} \right ) \; ,  \qquad
W^{D.(a)}_{r} = \pmatrix{0\cr0\cr0}    \; ,
$$
$$
W^{D.(a)}_{t} =
\left ( \begin{array}{c}
              0  \\
              0  \\
              rF(r)
\end{array} \right ) \; , \qquad
                          W^{D.(a)}_{\phi } =
                        \left ( \begin{array}{c}
-(r^{2}K + 1/e) \sin\theta \cos\phi \\
-(r^{2}K + 1/e) \sin\theta \sin\phi \\
        {1 \over e} (\cos\theta - 1)
\end{array} \right )   \; .
\eqno(A.4)
$$

\noindent It should be noticed that the~factor $(r^{2} K(r) + 1/ e)$  will
vanish when $K = -1 / e r^{2}$. Thus, only  the~delicate fitting of
the~single proportional   coefficient  (it  must  be
taken as $-1/ e)$  results in the~actual  formal  simplification  of
the~non-Abelian monopole potential.

There exists close connection between $W^{D.(a)}_{\phi }$  from (A.4) and
the~Dirac's expression for the~Abelian monopole potential
(supposing that $\vec{n} = (0, 0 ,-1)$):
$$
A^{D.}_{\alpha} = g \;\left ( \; 0 ,\; {{[\; \vec{n} \; \vec{r}\; ]} \over
{(r + \vec{r} \; \vec{n} )\; r }}\; \right ) \;\; , \qquad or \qquad
A^{D.}_{\phi } = - g \; ( \cos \theta  - 1 ) \; .
\eqno(A.5)
$$

\noindent So, $W^{triv.}_{(a)\alpha }(x)$  from (A.4)
 (produced by setting   $K = - 1 / e r^{2}$)
can be thought of  as  the~result of embedding the~ Abelian
potential (A.5) in the~non-Abelian gauge scheme:
$W^{(a)D.}_{\alpha }(x) \equiv  ( 0 , 0 , A^{D.}_{\alpha }(x))$.
The quantity $W^{(a)D.}_{\alpha }(x)$  labelled with symbol  $D.$
will be  named
after its Abelian counterpart; in other words, this potential will
be treated as relating to the Dirac's non-Abelian   gauge  in
the~isotopic space.

In Abelian  case,  the  Dirac's  potential $A^{D.}_{\alpha }(x)$
 can  be
converted into the~Schwinger form $A^{S.}_{\alpha }(x)$
$$
A^{S.}_{\alpha } = \left ( 0 ,\; g \; {{[\; \vec{r}\; \vec{n}\; ] \;
( \vec{r} \; \vec{n})} \over {(r^{2} \; - \; ( \vec{r}\; \vec{n})^{2})r}} \right ) \; ,
\qquad  or \qquad A^{S.}_{\phi } = g \; \cos \theta
\eqno(A.6)
$$

\noindent by means of the~following transformation
$$
A^{S.}_{\alpha }\; = \;A ^{D.}_{\alpha } \; +  \;
{{\hbar c} \over{ie}} \; S \; {{ \partial} \over { \partial x^{\alpha }}} \; S^{-1},
\qquad
S(x) = \exp  (-i{eg \over \hbar c} \phi  ) \;\; .
$$

\noindent It is  possible  to  draw  an~analogy  between  the~Abelian  and
non-Abelian   models.   That is, we    may introduce the~Schwinger
non-Abelian basis in the~isotopic space:
$$
( \Phi ^{D.(a)}, W^{D.(a)}_{\alpha }) \qquad \stackrel{\vec{c}\; '}
\Longrightarrow
 \qquad
 ( \Phi ^{S.(a)}, W^{S.(a)}_{\alpha } ) \; , \qquad
\vec{c}\; ' = ( 0 , 0 , - \tan  \phi /2 ) \;  ;
\eqno(A.7a)
$$

\noindent where
$$
O(\vec{c}~') = \pmatrix{ \cos \phi & \sin \phi & 0 \cr
                        -\sin \phi & \cos \phi & 0 \cr
                         0         &   0       & 1 }  \;  .
$$

\noindent Now an~explicit form of the~monopole potential is given by
$$
               W^{S.(a)}_{\theta} =
\left ( \begin{array}{c}
              0  \\( r^{2} K + 1/e ) \\ 0
\end{array} \right ) \; , \qquad
 W^{S.(a)}_{\phi } =
\left ( \begin{array}{c}
             -(r^{2} K + 1/e)  \\ 0  \\ {1\over e} \cos \theta
\end{array} \right )\; ,
$$
$$
W^{S.(a)}_{r} = \pmatrix{0\cr0\cr0}, \qquad
W^{S.(a)}_{t} = \pmatrix{ 0 \cr 0 \cr rF(r)} , \qquad
 \Phi ^{S.(a)} = \pmatrix{0 \cr 0 \cr r\Phi (r)}
\eqno(A.7b)
$$

\noindent where the symbol  $S.$  stands for the~Schwinger gauge.
Both $D.$- and $S.$-gauges (see (A.4) and (A.7b))  are  unitary
ones in the~isotopic space due to the~respective  scalar  fields
$\Phi ^{D.}_{(a)}(x)$   and $\Phi ^{S.}_{(a)}(x)$  are
 $x_{3}$-unidirectional,  but  one  of  them
(Schwinger's) seems simpler than another (Dirac's).

For the following it  will  be  convenient  to  determine
the~matrix $0(\vec{c}~'')$ relating the~Cartesian   gauge  of
isotopic space with Schwinger's:
$$
O(\vec{c}~'') = O(\vec{c}~') O(\vec{c}) =
\pmatrix{ \cos\theta  \cos\phi & \cos\theta \sin\phi & -\sin\theta \cr
         -\sin\phi             & \cos\phi            &   0         \cr
          \sin\theta  \cos\phi & \sin\theta \sin\phi &  \cos\theta } ,
$$
$$
\vec{c}~'' = (+ \tan\theta /2  \tan\phi /2, - \tan\theta /2, -\tan\phi /2).
\eqno(A.8)
$$

This matrix  $O(\vec{c}~'')$  is also well-known in
other  context  as  a~matrix  linking  Cartesian  and  spherical  tetrads
in  the space-time of special relativity  (as well as in a~curved  space-time
of spherical symmetry)
$$
x^{\alpha } = (x^{0}, x^{1}, x^{2}, x^{3}) \; , \;\;
dS^{2}= [(dx_{0})^{2} - (dx_{1})^{2} - (dx_{2})^{2} - (dx_{3})^{2}] \; , \;\;
e^{\alpha }_{(a)}(x) = \delta ^{\alpha }_{a}
\eqno(A.9a)
$$

\noindent and
$$
x'^{\alpha } = ( t , r , \theta  , \phi  ) \; , \qquad
dS^{2} = [dt^{2} - dr^{2} - r^{2}(d\theta ^{2} +
\sin ^{2} \theta  d\phi ^{2})] \; ,
$$
$$
e^{\alpha'}_{(0)} = ( 1 , 0 , 0 , 0 )\; ,  \qquad
e^{\alpha'}_{(1)} = ( 0, 0 , 1/r, 0 )\; ,
$$
$$
e^{\alpha'}_{(2)} = ( 0 ,0 , 0 ,1/ r \sin \theta)\; ,  \qquad
e^{\alpha'}_{(3)} = ( 0 , 1 , 0 , 0 ) \;\;  .
 \eqno(A.9b)
$$

Below we  review briefly   some   relevant  facts  about
the~tetrad formalism. In  the~presence of an~external  gravitational
field, the~starting Dirac equation
 $(i \gamma ^{a} \partial /\partial x^{a} - m ) \Psi (x) = 0$
is generalized into  [20-29]
$$
[\; i \gamma ^{\alpha}(x) \; (\partial_{\alpha} \; + \;
 \Gamma _{\alpha }(x) ) \;  - \;  m \; ] \; \Psi (x)  = 0
\eqno(A.10)
$$

\noindent where $\gamma ^{\alpha }(x) = \gamma ^{a} e^{\alpha }_{(a)}(x)$, and
$e^{\alpha }_{(a)}(x)$,
 $\Gamma _{\alpha }(x) = {1\over 2}
  \sigma ^{ab} e^{\beta }_{(a)} \nabla _{\alpha }(e^{\alpha }_{(b)\beta })$,
$\nabla _{\alpha }$ stand for a~tetrad, the~bispinor connection, and
the~covariant derivative symbol, respectively. In the~spinor basis:
$$
                      \psi (x)=
\left ( \begin{array}{c}
                  \xi (x) \\  \eta (x)
\end{array} \right ) ,   \qquad
\gamma ^{a} =
\left ( \begin{array}{cc}
                   0  & \bar{\sigma}^{a} \\
                   \sigma^{a}  &       0
\end{array} \right ) , \qquad
\sigma ^{a} = (I ,+ \sigma ^{k}), \qquad
\bar{\sigma }^{a}  = (I, -\sigma^{k})
$$

\noindent ($\sigma ^{k}$ are the two-row Pauli spin matrices; $k = 1,2,3$) we
 have  two equations
$$
i \sigma ^{\alpha }(x) \;[\; \partial_{\alpha} \; + \;
  \Sigma _{\alpha }(x)\; ] \; \xi (x) =  m \;\eta (x),  \qquad
i \bar{\sigma}^{\alpha }(x)\; [\; \partial_{\alpha } \; + \;
 \bar{\Sigma}_{\alpha }(x)\; ] \; \eta (x) = m \; \xi (x)
\eqno(A.11)
$$

\noindent where the~symbols
$\sigma ^{\alpha }(x), \bar{\sigma }^{\alpha }(x),
\Sigma _{\alpha }(x), \bar{\Sigma }_{\alpha }(x)$
 denote respectively
$$
\sigma ^{\alpha }(x)= \sigma ^{a} e^{\alpha }_{(a)}(x),\qquad
\bar{\sigma}^{\alpha }(x)= \bar{\sigma}^{a} e^{\alpha }_{(a)}(x),
$$
$$
\Sigma _{\alpha }(x) =
{1\over 2} \Sigma ^{ab} e^{\beta }_{(a)}
\nabla _{\alpha }(e_{(b)\beta }) ,   \qquad
\bar{\Sigma}_{\alpha }(x) = {1\over 2} \bar{\Sigma}^{ab} e^{\beta }_{(x)}
\nabla _{\alpha }(e_{(b)\beta }) ,
$$
$$
\Sigma ^{ab} = {1\over 4}(\bar{\sigma}^{a} \sigma^{b} -
                          \bar{\sigma}^{b} \sigma^{a}),     \qquad
\bar{\Sigma}^{ab} = {1\over 4} (\sigma^{a} \bar{\sigma}^{b} -
                                \sigma^{b} \bar{\sigma}^{a})  \; .
$$

The form of equations (A.10), (A.11)  implies  quite definite
their symmetry properties. It is common,  considering  the~Dirac
equation in the~same space-time, to use  some   different  tetrads
$e^{\beta }_{(a)}(x)$ and $e'^{\beta }_{(b)}(x)$, so that  we have
the~equation  (A.10)  and
an~analogous one with a~new tetrad mark.  In  other  words,  together
with  (A.10)  there  exists  an~equation  on $\Psi'(x)$, where
quantities $\gamma'^{\alpha }(x)$ and $\Gamma'_{\alpha}(x)$,
in contrast with $\gamma^{\alpha }(x)$ and $\Gamma_{\alpha}(x)$,
are based on a~new tetrad
$e'^{\beta }_{b)}(x)$ related  to
 $e^{\beta }_{(a)}(x)$  through a~certain local Lorentz matrix
$$
e'^{\beta }_{(b)}(x) \; =  \; L^{\;\;a}_{b}(x) \; e^{\beta }_{(a)}(x) \; .
\eqno(A.12a)
$$

\noindent It may be shown that these two Dirac equations on
 functions $\Psi (x)$ and $\Psi'(x)$
 are related to each other  by  a~definite  bispinor transformation:
$$
\xi'(x)  = B(k(x)) \xi (x), \qquad
\eta'(x) = B^{+}(\bar{k}(x)) \eta (x) \; .
\eqno(A.12b)
$$

\noindent Here, $B(k(x)) = \sigma ^{a} k_{a}(x)$ is a~local matrix
 from the $SL(A.C)$ group;
4-vector $k_{a}$ is the well-known parameter on this group
([67]; also see in [64]).
The~matrix $L^{\;\;a}_{b}(x)$ from (A.12a)  may  be  expressed as a~function of
arguments $k_{a}(x)$  and $k^{*}_{a}(x)$ :
$$
L^{a}_{b}(k, k^{*})  \; = \; \bar{\delta}^{c}_{b} \;
[\; - \delta^{a}_{c} \; k^{n} \; k^{*}_{n} \; + \;  k_{c} \; k^{a*}\;  +  \;
  k^{*}_{c} \; k^{a} \; + \; i \epsilon ^{\;\;anm}_{c}\; k_{n}\; k^{*}_{m} \;]
\eqno(A.12c)
$$

\noindent where  $\bar{\delta}^{c}_{b}$  is a~special Cronecker symbol
$$
\bar{\delta}^{c}_{b} = 0 \;\;\; if \;\;\;c \neq \; ; \;\;= +1 \;\;\; if \;\; \;
 c = b = 0 \; ; \;\; = -1 \;\;\; if \;\; \;  c = b = 1,2,3 \;\; .
$$

It is normal practice that some different  tetrads  are
used in examining the~Dirac equation on the~same
Rimannian space-time background. If there is a~need to analyze some
correlation between  solutions in those distinct tetrads,  then  it
is important  to know what are the~relevant  gauge
transformations over the~spinor wave functions.
In particular, the~matrix
relating  spinor  wave  functions  in   Cartesian   and  spherical
tetrads (see (A.9)) is as follows
$$
B = \pm \pmatrix{ \cos\theta /2  \; e^{i\phi  /2} &
                   \sin\theta /2 \; e^{-i\phi /2} \cr
                 -\sin\theta /2  \; e^{i\phi  /2} &
                  \cos\theta /2  \; e^{-i\phi /2} }
\equiv  B( \vec{c}~'') =
\pm {{ I  - i \vec{\sigma } \vec{c}~'' } \over
{ \sqrt{1 -  (\vec{c}~'')^{2} }}}   \; .
\eqno(A.12d)
$$

\noindent The vector matrix $L^{\;\;a}_{b}(\theta ,\phi )$ referring to
the spinor's  $B(\theta ,\phi )$
is the same as $O(\vec{c}~'')$  from (A.8). It is significant that the~two
gauge transformations, arising in quite different contexts, correspond so
 closely  with each other.

This basis of spherical tetrad  played  a~substantial role in the present work.
This Schr\"odinger frame of spherical tetrad [29] was  used  with  great  efficiency  by
Pauli [30] when investigating the~problem of  allowed  spherically
symmetrical wave  functions  in  quantum  mechanics. Below, we briefly review
some results of this investigation.
Let the $J^{\lambda }_{i}$   denote
$$
J_{1}= (\; l_{1} + \lambda\; {{\cos  \phi  }\over {\sin \theta}}\; ),\qquad
J_{2}= (\; l_{2} + \lambda\; {{\sin  \phi  }\over {\sin \theta}}\; ),\qquad J_{3} = l_{3} \; .
$$

\noindent At an~arbitrary $\lambda$, as readily  verified, those $J_{i}$
 satisfy the~commutation rules of the~Lie algebra
$SU(2): [ J_{a},\; J_{b} ] = i \; \epsilon _{abc} \; J_{c}$.
As known,  all  irreducible  representations  of  such  an~ abstract
algebra are determined by a~set of weights
   $j = 0, 1/2, 1, 3/2,... \; \; ({\em dim} \;j = 2j + 1)$.
Given the~explicit expressions of  $J_{a}$   above,  we  will
find functions
               $\Phi ^{\lambda }_{jm}(\theta ,\phi )$
on which the~representation of  weight $j$ is realized. In agreement with
the~generally known method, those solutions are to be established by
the~following relations
$$
J_{+} \; \Phi ^{\lambda }_{jj} \; = \; 0 \;\; , \qquad
\Phi ^{\lambda }_{jm} \; = \; \sqrt{{(j+m)! \over (j-m)! \; (2j)! }} \; J^{(j-m)}_{-} \;
\Phi^{\lambda}_{jj} \;\; ,
\eqno(A.13)
$$
$$
J_ {\pm} \; = \; ( J_{1} \pm i J_{2}) \; = \;
e^{\pm i\phi }\; [\; \pm { \partial \over \partial \theta } \; + \;
 i \cot \theta \; { \partial \over  \partial \phi} \; + \;
 { \lambda \over  \sin  \theta }\; ] \;\; .
$$

\noindent From the equations
$J_{+} \; \Phi ^{\lambda }_{jj} \; = \; 0 \;$ and
$\; J_{3} \; \Phi ^{\lambda }_{jj} \;  = \; j \; \Phi ^{\lambda }_{jj}$,
it follows that
$$
\Phi ^{\lambda }_{jj}  =
N^{\lambda }_{jj} \;  e^{ij\phi} \; \sin^{j}\theta \;\;
{( 1 + \cos \theta  )^{+\lambda /2} \over ( 1 - \cos \theta )^{\lambda /2}} ,\;
N^{\lambda }_{jj}  = {1 \over \sqrt{2\pi}} \; { 1 \over 2^{j} } \;
 \sqrt{{(2j+1) \over \Gamma(j+m+1) \; \Gamma(j-m+1)}} \; .
$$

\noindent Further, employing (A.13) we produce the~functions
$\Phi ^{\lambda }_{jm}$
$$
\Phi ^{\lambda }_{jm} \; = \; N^{\lambda }_{jm} \;  e^{im\phi} \;
 {1 \over \sin^{m}\theta } \;\;
{(1 - \cos \theta)^{\lambda/2} \over (1 + \cos \theta)^{+\lambda/2}} \; \times
$$
$$
({ d \over d \cos  \theta})^{j-m} \; [\; (1 + \cos  \theta ) ^{j + \lambda } \;
(1 - \cos  \theta ) ^{j-\lambda } \; ]
\eqno(A.14)
$$

\noindent where
$$
N^{\lambda }_{jm} \;  = \; {1 \over \sqrt{2\pi} 2^{j}} \;
 \sqrt{{(2j+1) \; (j+m)! \over
2(j-m)!  \Gamma(j + \lambda +1) \; \Gamma(j- \lambda +1)}} \;\; .
$$

\noindent The Pauli criterion tells us that the $(2j + 1)$ functions
$ \Phi ^{\lambda }_{jm}(\theta ,\phi ), \; m = - j,... , +j$
so  constructed, are  guaranteed  to be a~basis for a~finite-dimension
representation, providing that the~functions
$\Phi ^{\lambda }_{j,-j}(\theta ,\phi )$,
found by this procedure,  obey the~identity
$$
J_{-} \;\; \Phi ^{\lambda }_{j,-j} \; = \; 0 \; .
\eqno(A.15a)
$$

\noindent After substituting the~function
 $\Phi ^{\lambda }_{j,-j}(\theta ,\phi )$, the~relation (A.15a)  reads
$$
J_{-} \; \Phi ^{\lambda }_{j,-j} \; = \;  N^{\lambda }_{j,-j} \;
e^{-i(j+1)\phi }\; (\sin \theta)^{j+1}  \;
{(1 - \cos  \theta )^{\lambda /2} \over (1 + \cos  \theta )^{\lambda /2}} \; \times
$$
$$
({d \over d \cos \theta})^{2j+1} \; [\; (1 + \cos  \theta )^{j+\lambda } \;
(1 - \cos  \theta )^{j-\lambda } ) \; ]\;  = \;  0
\eqno(A.15b)
$$

\noindent which in turn gives the~following restriction on
 $j$  and $\lambda $
$$
({d \over d \cos  \theta})^{2j+1} \; [\; (1 + \cos  \theta  )^{j+\lambda } \;
(1 - \cos  \theta  )^{j-\lambda } \; ] \;  = \; 0 \; \; .
\eqno(A.15c)
$$

\noindent But the~relation (A.15c) can be satisfied  only  if  the~factor
 $P(\theta )$,
subjected to the~operation of taking derivative
$( d/d \cos \theta ) ^{2j+1}$,
is a~polynomial of degree $2j$  in $ \cos \theta$. So, we have (as a~result
of the~Pauli criterion)

\vspace{5mm}
1.  {\em the} $\lambda$ {\em is allowed to take values}
$, +1/2,\; -1/2,\; +1,\; -1, \ldots$
\vspace{5mm}

\noindent Besides, as the~latter condition is satisfied,
$P(\theta )$  takes different forms depending on
the $(j , \lambda)$-correlation:
$$
P(\theta ) \; = \; (1 + \cos \theta )^{j+\lambda } \;
  (1 - \cos \theta )^{j - \lambda } \; = \;
P^{2j}(\cos \theta ),\qquad if\qquad j = \mid \lambda \mid,
\mid \lambda \mid +1,...
$$
or
$$
P(\theta ) \; = \;
{ P^{2j+1}(\cos \theta ) \over \sin \theta }, \qquad if \qquad
 j = \mid \lambda \mid +1/2, \mid \lambda \mid +3/2,...
$$

\noindent so that the second necessary condition  resulting from
the~Pauli criterion  is

\vspace{5mm}
2.  {\em given } $\lambda$ {\em according to 1.,
  the number j is allowed to take values}
  $j = \mid \lambda \mid, \mid \lambda \mid +1,...$
\vspace{5mm}

\noindent We draw attention to   that
the  Pauli  criterion
$
J_{-} \Phi _{j,-j}(t,r,\theta ,\phi )\; =\; 0
$
affords the~condition that is invariant relative to possible  gauge
transformations. The function $\Phi _{j,m}(t,r,\theta ,\phi )$
may be subjected  to any gauge transformation. But if  all the~components
 $J_{i}$ vary  in a~corresponding way too, then the
Pauli  condition provides the~same result on $(j,\lambda)$-quantization.
In contrast
to this, the common  requirement to be a~single-valued function of spatial
points, often applied to produce a~criterion on selection  of
allowable wave functions in quantum mechanics, is  not
invariant under gauge transformations and can easily be destroyed
by a~suitable gauge one.
Also, it should be noted  that  the~angular  variable $\phi $  is  not
affected (charged) by the Pauli criterion; instead,
a~variable that  works above is the~$\theta$. Significantly,  in
the~contrast to this, the~well-known procedure of deriving
 the~electric charge quantization  condition  from investigating continuity
properties of quantum mechanical wave functions,  such
a~working variable is the $\phi $.

If the~first and second Pauli consequences fail, then we face
rather unpleasant mathema\-ti\-cal and physical problems (Reader  is
referred  to  the~Pauli article [30] for more detail about
those peculiarities). As a~simple illustration,  we  may  indicate
the~familiar case  when $\lambda= 0$; if
 the~second  Pauli  condition is violated, then we will have the~integer and
half-integer  values  of the~orbital angular momentum number
$l = 0, 1/2, 1, 3/2,\ldots\;$
As regards  the~Dirac  electron  with  the  components  of  the  total
angular momentum in the~form (1.2),
we have to employ the~above Pauli  criterion in  the~constituent
form owing to $\lambda $  changed into $\Sigma _{3}$.
Ultimately, we obtain the~allowable set $J = 1/2, 3/2, \ldots$.

A~fact of primary  practical importance to  us  is  that  the~functions
$\Phi ^{\lambda }_{jm}(\theta, \phi )$ constructed  above  relate
directly  to  the~known Wigner $D$-functions [18]:
$
\Phi ^{\lambda }_{jm}(\theta , \phi ) \; =  \;
(-1)^{j-m} \; D^{j}_{-m, \lambda}(\phi, \theta, 0)
$.

\subsection*{Supplement B. On complete set of operators.}

For our purposes we may take a Dirac's generalized operator; in spherical
tetrad and Scwinger isotopic basis it is
$$
\hat{K} = \;  + i \; \gamma ^{0} \; \gamma ^{3}\;  \Sigma _{\theta ,\phi } =
i \; \gamma ^{0} \; \gamma ^{3}
\;\left [\; i \gamma ^{1}\; \partial _{\theta } \; + \;
\gamma ^{2}\;  { i \partial _{\phi } + (i\sigma ^{12} + t^{3}) \cos \theta
\over  \sin \theta } \; \right ] \; .
\eqno(B.1)
$$

\noindent Now, taking into account already known action  of the
$\Sigma _{\theta ,\phi }$ on wave functions with fixed numbers  $j$ ¨ $m$,
for $\hat{K} \; \Psi _{jm\delta }(x)$ we get
$$
\hat{K} \; \Psi ^{A(B)}_{jm\delta } (x)  = {e^{-i\epsilon t} \over r} \times
$$
$$
\left [ \; b \;  T_{+1} \otimes  \left  ( \begin{array}{c}
f_{4}\; D_{-3/2} \\ f_{3}\; D_{-1/2} \\
f_{2}\; D_{-3/2}   \\ f_{1}\; D_{-1/2}
\end{array} \right )     \;   +   \;
a \; e^{iB} \; T_{0} \otimes
\left ( \begin{array}{c}
\delta  \; h_{1} \; D_{-1/2} \\ \delta  \; h_{2} \; D_{+1/2} \\
h_{2}\; D_{-1/2}  \\   h_{1}\; D_{+1/2}
\end{array} \right )  \; + \; \delta  \; e^{iA} \; b \;
T_{-1} \otimes  \left ( \begin{array}{c}
f_{1} \; D_{+1/2} \\ f_{2} \; D_{+3/2} \\
f_{3} \; D_{+1/2} \\    f_{4} \; D_{+3/2}
\end{array} \right )  \; \right ]    \; .
$$

\noindent Correspondingly, from the equation  $\hat{K}\;
\Psi ^{A(B)}_{jm\delta }(x) = \lambda \; \Psi ^{A(B)}_{jm\delta }(x)$
it follows
$$
\left. \begin{array}{llll}
  b \; f_{4} = \lambda  \; f_{1} \; , \qquad
& b \; f_{1} = \lambda  \; f_{4} \; , \qquad
& b \; f_{3} = \lambda  \; f_{2} \; , \qquad
& b \; f_{2} = \lambda  \; f_{3} \; ,   \\
\delta  \; a \; h_{1} = \lambda  \; h_{1} \; , \;\;\;
& a \; h_{1} = \delta  \; \lambda  \; h_{1} \; , \;\;\;
& \delta \;  a  \; h_{2} = \lambda  \; h_{2}\; , \;\;\;
& a \;h_{2} = \delta  \; \lambda  \; h_{2}  \; .
\end{array} \right.
\eqno(B.2)
$$

\noindent
Let us suppose that all radial functions $f_{1},\ldots ,\;f_{4}$, and
$h_{1}, \;h_{2}$ are not null ones; then from (B.2) we arrive at two
non-consistent with each other relations:
$\lambda ^{2} = b^{2}$   and   $\lambda ^{2} = a^{2}$,   but $a$  and $b$
are already given as basically different:
$ a  = (j  + 1/2) \; , \; \; b = \sqrt{(j - 1/2)(j + 3/2)} \;$.
Therefore, the equations in (B.2) may be satisfied
only as follows:

{\bf 1:}
$$
f_{i}= 0 \; ; \; \; h_{1} \; , \; \; h_{2} \neq  0 \qquad
\Longrightarrow \;\;\;
\lambda ^{(1)} = \delta  \; a \;\;\; (\delta = \pm  1) \; ,
\eqno(B.3a)
$$

\noindent and  proper wave functions of $\hat{K}$ are
$$
\Psi ^{(B)}_{jm\delta (\delta )}(x) = e^{iB} \; T_{0} \otimes
\left ( \begin{array}{c}
h_{1} \;  D_{-1/2}   \\  h_{2} \; D_{+1/2}   \\
\delta  \; h_{2} \; D_{-1/2}  \\    \delta  \; h_{1} \; D_{+1/2}
\end{array} \right )  \; ;
\eqno(B.3b)
$$

\noindent corresponding  system of radial equations is
$$
\epsilon \; h_{2} -i{d \over dr}\;  h_{2} \
- \; {ia \over r} \; h_{1} \; - \; \delta  \; m\; h_{1} = 0   \; , \;\;
$$
$$
\epsilon \; h_{1}\; + \; i {d \over dr} \; h_{1} \; + \; {ia \over r}\;
h_{2} \; -  \; \delta  \; m \; h_{2} = 0     \; .
\eqno(B.3c)
$$

{\bf 2:}

$$
f_{i} \neq  0 \;\; ;  \;\;\; h_{1}, \;\;\; h_{2} = 0
 \qquad \Longrightarrow \;\;\;
\lambda ^{(2)} = \mu  \; b \;\;\;
(\mu  = \pm  1)     \; ,
\eqno(B.4a)
$$

\noindent and proper wave functions of the $\hat{K}$  are
($f_{4} = \mu  \; f_{1}, \; f_{3} = \mu  \; f_{2}$)
$$
\Psi ^{A}_{jm\delta \mu } (x) = {e^{-i\epsilon t} \over r} \;
\left [ T_{+1} \otimes
\left ( \begin{array}{c}
f_{1} \; D_{-3/2}         \\          f_{2} \; D_{-1/2} \\
\mu \; f_{2} \;  D_{-3/2} \\  \mu \; f_{1} \; D_{-1/2}
\end{array} \right )  \; + \; \delta  \; \mu  \; e^{iA} \; T_{-1} \otimes
\left ( \begin{array}{c}
f_{1} \; D_{+1/2} \\ f_{2} \; D_{ +3/2} \\
\mu  \; f_{2} \; D_{+1/2} \\ \mu \; f_{1}  \; D_{+3/2}
\end{array} \right ) \right ]  \;
\eqno(B.4b)
$$

\noindent the radial system will be as follows
$$
\epsilon  \; f_{2} - i {d \over dr}\; f_{2}-
{ib \over r} \; f_{1} \; - \; \mu  \; m \; f_{1} = 0 \; , \;\;
$$
$$
\epsilon \; f_{1} \;+\; i {d \over dr} \; f_{1} \; + \; {ib \over r} \;
f_{2} \; - \mu  \; m  \;f_{2} = 0   \; .
\eqno(B.4c)
$$

\noindent Thus, the complete set of operators
$ \hat{H} , \; \vec{J}^{2} , \; J_{3} , \; \hat{N}_{A} , \; \hat{K}$
leads us to  basis functions of two different types:
$$
\Psi ^{(B)}_{jm\delta (\delta )}(x) \; ,\;\; \lambda ^{(1)} =
\delta  \; a   \;\;\; and   \qquad ¨ \qquad
\Psi ^{(A)}_{jm\delta \mu }(x) \; , \;\; \lambda ^{(2)} = \mu \;b\; ;
\eqno(B.5)
$$

\noindent that is every state with fixed $\epsilon ,\; j,\; m, \; \delta $
 is yet triply degenerated; the situation may illustrated by
the scheme
$$
\Psi ^{A(B)}_{\epsilon jm\delta}(x) \qquad \Longrightarrow  \qquad
\left \{ \; \Psi ^{A} _{\epsilon jm\delta (\delta )}(x)
\oplus \Psi ^{B}_{\epsilon jm\delta \mu}(x) \; \right \} \; .
\eqno(B.6)
$$

\noindent Now let us consider the case of minimal $j$:
$$
\Psi ^{min.}_{\delta } (t,r) = {e^{- i \epsilon t} \over r} \times
$$
$$
\left [ T_{+1} \otimes       \left ( \begin{array}{c}
 0     \\       f_{2} \;  D_{-1/2}   \\
 0     \\          f_{4}  \; D_{-1/2}
\end{array} \right ) \;  + \;
e^{iB} \; T_{0} \otimes       \left ( \begin{array}{c}
h_{1} \; D_{-1/2}   \\   h_{2} \; D_{+1/2}   \\
\delta \; h_{2} \; D_{-1/2}   \\     \delta \; h_{1} \; D_{+1/2}
\end{array}\right ) \;  +
\; e^{iA} \; \delta \; T_{-1} \otimes
\left ( \begin{array}{c}
f_{4} \; D_{+1/2}  \\   0   \\   f_{2} \; D_{+1/2}  \\   0
\end{array} \right ) \right ] \; .
$$

\noindent From the equation
$\hat{K}  \Psi ^{min.}_{\delta }  = \lambda \; \Psi ^{min.}_{\delta }$
we can get two proper values $\lambda$:
$$
\lambda = \delta \; : \qquad
\Psi ^{min.}_{\delta (\lambda = \delta)} (t,r) =
e^{iB} \; T_{0} \otimes       \left ( \begin{array}{c}
h_{1} \; D_{-1/2}   \\   h_{2} \; D_{+1/2}   \\
\delta \; h_{2} \; D_{-1/2}   \\  \delta \;    h_{1} \;D_{+1/2}
\end{array}\right )  \; ;
\eqno(B.7a)
$$
$$
\lambda = 0 \; : \;\;\;
\Psi ^{min.}_{\delta (\lambda = 0 )} (t,r) =
\left [
T_{+1} \otimes       \left ( \begin{array}{c}
 0     \\       f_{2} \;  D_{-1/2}   \\
 0     \\          f_{4}  \; D_{-1/2}
\end{array} \right ) \;  + \;
\; e^{iA} \; \delta \; T_{-1} \otimes
\left ( \begin{array}{c}
f_{4} \; D_{+1/2}  \\   0   \\   f_{2} \; D_{+1/2}  \\   0
\end{array} \right )  \;\right ] \; .
\eqno(B.7b)
$$

\end{document}